\documentclass[journal]{IEEEtran}

\usepackage{graphicx}
\usepackage{booktabs}
\usepackage{multirow}
\usepackage{amsmath}
\usepackage{amssymb}
\usepackage{xcolor}
\usepackage{tikz}
\usetikzlibrary{shapes.geometric, arrows.meta, positioning, calc, fit, backgrounds, patterns, patterns.meta}
\usepackage{url}
\usepackage{xurl}   
\usepackage{array}
\usepackage{tabularx}
\usepackage{makecell}
\usepackage{enumitem}

\definecolor{darkgreen}{RGB}{0,120,50}
\definecolor{darkblue}{RGB}{0,70,150}
\definecolor{darkorange}{RGB}{200,100,0}
\definecolor{p1c}{HTML}{DCE8F5}
\definecolor{p2c}{HTML}{BBD3EC}
\definecolor{p3c}{HTML}{A9D8D1}
\definecolor{p4c}{HTML}{BFE1C6}
\definecolor{p5c}{HTML}{ECD3A9}
\definecolor{p6c}{HTML}{DCCBB1}
\definecolor{edgec}{HTML}{53616F}
\definecolor{gridc}{HTML}{D2D9E0}
\definecolor{txtc}{HTML}{263039}
\definecolor{dimc}{HTML}{7C8894}
\definecolor{starc}{HTML}{8E3323}
\definecolor{bandc}{HTML}{ECECEC}

\newcommand{\reg}[1]{\texttt{#1}}

\title{AI-Assisted Design of a Post-Quantum Cryptographic
       Accelerator: A Deployed-Silicon Case Study}

\author{Jungmin~Park, Eunha~Kim, Wooseop~Kim, Seongjoon~Cho, and
        Byungho~Cha
        \thanks{J.~Park and B.~Cha are with Lucid Motors, Newark, CA, USA.
                E-mail: jungminpark@lucidmotors.com;
                byunghocha@lucidmotors.com.}%
        \thanks{E.~Kim, W.~Kim, and S.~Cho are with EYL Inc., Seoul,
                Republic of Korea.  E-mail: ehkim@eylpartners.com;
                wskim@eylpartners.com; jhcho@eylpartners.com.}}

\begin{document}


\maketitle

\begin{abstract}
Post-quantum migration is mandated on published timelines, and silicon
that ships with a defect cannot be patched remotely.  The standard acceptance
gate cannot detect an entire class of ML-DSA defects.  Signing resamples until a candidate meets its norm bounds, so
the executed path varies with the message, whereas known-answer tests
(KATs) sample fixed values and reach only the depths their seeds
trigger.  Our accelerator passed its full KAT regression while carrying a
norm check that outran block-RAM latency, leaving each candidate's
final coefficients unverified; the escape surfaced at reject-loop
iteration~5.  The blind spot lies in the instrument, not the
engineer; care cannot remove it.  We replace that gate.  A byte-exact golden-reference
oracle paired with randomized adversarial soak drives the rejection
loop past any fixed vector, closing the gap: 301,343
data-dependent signings, zero escapes.  Because the gate judges
artifacts and never authors, trust becomes separable from authorship,
making AI authorship an answerable question.  We report 232 logged
experiments in which an agentic large language model drove a unified
ML-KEM-768 and ML-DSA-65 accelerator with on-chip key custody from RTL
to PCIe bring-up on one Kintex-7 XC7K160T, shipped at 98.5\% slice
occupancy.
Success was 71.6\%, following a hardware-coupling gradient, 77--85\%
for documentation and research against 50--53\% for synthesis and
bring-up, which observability can explain: failure concentrates
where corrective signals are physical-side only.  That so unreliable an
author produced an artifact byte-exact across all six FIPS
operations---its deployed baseline surviving the same 779,945-check
zero-failure soak---is the claim.
\end{abstract}

\begin{IEEEkeywords}
Post-quantum cryptography, FPGA, LLM-assisted design, design methodology,
empirical study, hardware security module, ML-KEM, ML-DSA, NTT, key custody,
agentic AI, design automation
\end{IEEEkeywords}

\section{Introduction}
\label{sec:intro}

Post-quantum cryptographic hardware is essential yet error-prone, and
the test that decides whether it may ship is blind to an entire class
of its defects.
NIST closed PQC standardization in August 2024 with FIPS~203
(ML-KEM~\cite{fips203,kyber2018}) and FIPS~204
(ML-DSA~\cite{fips204,dilithium2018}).  CNSA~2.0~\cite{cnsa20} and
NIST~IR~8547~\cite{nistir8547} attach dates, the latter proposing
deprecation of 112-bit-security RSA and ECC by 2030 and disallowance
of all quantum-vulnerable algorithms by 2035.
Migration is therefore a scheduled compliance obligation, and it lands
hardest on hardware.  Roots of trust,
secure elements, and signing devices stay in the field for a decade or
more and cannot be patched remotely.  A defect shipped in silicon is a
recall, not a release.

The defect class that concerns us is structural: ML-DSA signing is
rejection-sampled.  The signer resamples until the
candidate satisfies its norm bounds, so the number of iterations, and
with it the control path exercised, depends on the message.
Correctness is therefore a property of execution \emph{paths}.
Known-answer tests sample \emph{values} instead, and a fixed vector set
reaches only the reject-loop depths its own seeds happen to trigger.
Enlarging that set does not help: the coverage limit follows from how
the vectors are built, not from how many there are.

Our accelerator passed its full KAT
regression while carrying BUG-IDE-037: the rejection-sampling norm
check iterated over all 256 coefficients of each polynomial, but
block-RAM read latency delivered the final one or two \emph{after} the
accept/reject decision (Section~\ref{subsec:bugide037}).
The defect was reachable only at reject-loop
iteration~5,\footnote{BUG-IDE-\textit{NNN} labels are the project's
stable defect-log identifiers.  Iterations count signing resampling
rounds: the fifth candidate was the first whose only out-of-bound
coefficient fell in the unchecked tail, so the device accepted what a
correct signer rejects (first correct accept: iteration~8).} for one
message class, and every affected signature failed verification.  It
is one of eight
defect classes we catalogue, each caught by a validation layer the
others miss (Table~\ref{tab:bugtax}).

The gap lives in the instrument: a careful team using KAT as its
acceptance gate ships the same defect.  We therefore replace the gate
with a byte-exact golden-reference oracle paired with randomized
adversarial soak that drives the rejection loop past any fixed
vector's reach.
Because the replacement judges artifacts and never authors, it is
\emph{author-agnostic}: who wrote the RTL becomes a separate question.

That separation is what makes the AI question worth asking.  The
deadline is fixed, engineers fluent in both lattice cryptography and
FPGA timing closure are scarce, and agentic LLM coding assistants are
already entering hardware flows.  The open question is not whether they
will be used on PQC silicon but \emph{where in the flow they can be
trusted}.  The LLM-for-RTL
literature evaluates snippets and single modules by pass@$k$ at
simulation or lint
\cite{thakur2023benchmark,liu2023verilogeval,liu2025rtlcoder}, and even
agentic, tool-in-the-loop flows
\cite{thakur2023autochip,tsai2024rtlfixer,ho2024verilogcoder,he2024chateda}
stop short of validated, hardened operation on deployed silicon.  No
published study reports how assistant reliability varies \emph{across
the phases} of a real campaign.

Across the 2026 development cycle (2026-01 through 2026-06) an agentic
LLM assistant drove design, verification, synthesis, host-software development,
silicon bring-up, and security hardening of a unified ML-KEM-768 +
ML-DSA-65 accelerator with on-chip key custody, deployed on an
XC7K160T Kintex-7 over PCIe Gen2~$\times$8.  All 232 experiments were
logged.  ``Silicon'' herein means the physically deployed FPGA in live
operation, not simulation.

Three constraints made that campaign adversarial.
\emph{Area} was binding.  Both NIST Level-3 schemes plus the custody
HSM had to fit one mid-range XC7K160T, which forced a serialized
number-theoretic transform (NTT) and still left the shipped build at
98.5\% slice occupancy.  The limit was placement density rather than
arithmetic (Sections~\ref{subsec:resources} and~\ref{subsec:wrongepisodes}).
\emph{Timing} was tight.  The 500\,MHz domain of the Xilinx XDMA PCIe bridge sets the design's
worst-case setup slack at $+0.049$\,ns (0 failing among 178,247
endpoints design-wide); the path tipped negative during bring-up and
was recovered by a seed sweep because the cause was congestion, not logic.  \emph{Observability} was the worst of the
three.  A stale command register survived soft-reset,
wedging DMA ${\sim}$1-in-20 yet staying hidden for weeks because the
soak harness self-healed it.  That did not stop the assistant from
reaching a confident, plausible, and incorrect diagnosis in ten logged
episodes (Section~\ref{subsec:wrongepisodes}).

Where the corrective
signal for a task sits in an artifact the assistant can read, it
performs at software-domain rates.  Where that signal exists only on
the physical side of the boundary, in DMA-engine residue, reset-domain
asymmetry, placement density, or silicon-versus-simulator semantics, no
reasoning over readable artifacts recovers it, and reliability falls to
roughly a coin flip.  This \emph{hardware-coupling gradient} is a claim about
\emph{observability}, not that hardware is harder; its remedy is to make
the physical side readable.

This work makes three contributions:
\begin{enumerate}[label=\textbf{(C\arabic*)}]

\item \textbf{A validation construction that closes a gap KAT cannot
reach.}
\label{contrib:validation}
Fixed-vector testing is structurally incomplete for rejection-sampled
signature hardware; we demonstrate the consequence on deployed silicon
(BUG-IDE-037, passed by a full KAT regression) and close it with the oracle-plus-soak gate above, over 301,343 data-dependent signings with zero escapes.  The
construction
applies to any ML-DSA hardware effort; we generalize it into nine honesty
constraints under which trust derives from validation artifacts, never
from authorship (Section~\ref{sec:threats}).

\item \textbf{The hardware-coupling gradient and its cause.}
\label{contrib:empirical}
From 232 structured logs (40 task categories, 76 milestones) we show
that LLM success declines as tasks approach FPGA placement, timing, and
live-hardware state (77--85\% documentation and research; 50--53\%
synthesis and bring-up), and we identify the mechanism: failure
concentrates where corrective information is unobservable in any
artifact the model can read.  We also report an eight-class bug taxonomy and ten
``confidently wrong, then corrected'' episodes (Sections~\ref{subsec:bugtaxonomy} and~\ref{subsec:wrongepisodes}).

\item \textbf{Silicon existence proof under binding constraints.}
\label{contrib:artifact}
The loop produced a unified ML-KEM-768 + ML-DSA-65 accelerator plus
key-custody HSM co-resident on one XC7K160T at 98.5\% slice occupancy
with 49\,ps of setup margin at 500\,MHz and hardware-RNG-seeded keys under a device-bound
key-encryption key (KEK)---a combination that is, to our knowledge,
unique among unified PQC accelerators.  The artifact is byte-exact
across all six FIPS operations, survived a zero-failure 779,945-check
soak on the deployed firmware baseline v88 (v$N$ labels
successive firmware builds), and was productized through two releases to the final
FIPS-203/204 wire formats (Sections~\ref{sec:artifact}--\ref{sec:pcie}).
That an author measured at 71.6\% task success produced an artifact
with zero validation escapes is the concrete demonstration
of~\ref{contrib:validation}.

\end{enumerate}

\emph{Efficiency} here means consolidation rather than speed: both
schemes plus the key-custody hardware security module (HSM) share one
runtime-modulus-switched NTT/INTT datapath and one Keccak-f[1600]
core.
\emph{Security} means key custody: private keys are generated,
wrapped, and used on-die, device-bound and adversarially validated.  The claim is device binding,
not at-rest secrecy and not side-channel resistance.
Table~\ref{tab:comparison} places the artifact against the closest
unified ML-KEM$+$ML-DSA accelerators on exactly that basis.

Two 2026 works combine LLMs with PQC hardware,
LLM4PQC~\cite{perera2026llm4pqc} and Liao et
al.~\cite{liao2026llm4pqc}, the latter the first quantitative
LLM-versus-HLS comparison (up to $2.6\times$ kernel speedup); the
delta is one of depth rather than kind, contrasted axis by axis in
Table~\ref{tab:pqcllm} (Section~\ref{subsec:related_llm4pqc}).

The paper proceeds from background (Section~\ref{sec:background})
through the methodology (Section~\ref{sec:methodology}), the
empirical record (Section~\ref{sec:empirical}), and the artifact with
its PCIe system (Sections~\ref{sec:artifact}--\ref{sec:pcie}) to
related work (Section~\ref{sec:related}), threats to validity
(Section~\ref{sec:threats}), and the conclusion
(Section~\ref{sec:conclusion}).

\section{Background}
\label{sec:background}

\subsection{Preliminaries}
\label{subsec:mlkem}

ML-KEM-768 (FIPS~203) is a Module-LWE (MLWE) key-encapsulation mechanism
with rank $k = 3$ over $\mathcal{R}_q = \mathbb{Z}_q[x]/(x^n + 1)$,
$n = 256$, $q = 3329$, $\zeta = 17$.  Hashing uses SHAKE128/256 and
SHA3-256/512 (FIPS~202~\cite{fips202}).  Wire sizes are encapsulation key 1184~B,
decapsulation key 2400~B, ciphertext 1088~B, and shared secret 32~B.

ML-DSA-65 (FIPS~204) is a Fiat--Shamir-with-aborts lattice signature
(MLWE and Module-SIS problems) with $n = 256$,
$q = 8\,380\,417$, $\zeta = 1753$, $(k, \ell) = (6, 5)$, SHAKE256 mask
expansion, and rejection-sampling bounds $\gamma_1 = 2^{19}$,
$\gamma_2 = (q{-}1)/32$, $\beta = 196$, $\omega = 55$.  Wire sizes are
public key 1952~B, secret key 4032~B, and signature 3309~B.

\textbf{Shared-primitive opportunity.}
Both schemes share $\mathbb{Z}_q[x]/(x^n+1)$, $n = 256$: the NTT butterfly is identical in form across both moduli, and both
rely on the same Keccak permutation, so resource sharing is natural
and both schemes fit together on a Kintex-7 XC7K160T
(Section~\ref{subsec:resources}).

\subsection{LLM-Assisted Hardware Design: A Brief Primer}
\label{subsec:llmprimer}

LLMs can generate syntactically valid
Verilog~\cite{thakur2023benchmark,liu2023verilogeval,liu2025rtlcoder},
close simulation-feedback loops~\cite{thakur2023autochip,tsai2024rtlfixer,ho2024verilogcoder,xu2024meic},
and assist verification and security-sensitive RTL
repair~\cite{fang2025assertllm,ahmad2024hwsecbug}
(surveyed in Section~\ref{sec:related}).  To our knowledge, no prior work
closes the full path (RTL generation, synthesis, host software,
live-hardware bring-up, iterative hardening) on a complete
cryptographic system while reporting a campaign-level longitudinal
dataset of LLM-task outcomes.

\section{Methodology: An Agentic LLM-Assisted Design Loop}
\label{sec:methodology}

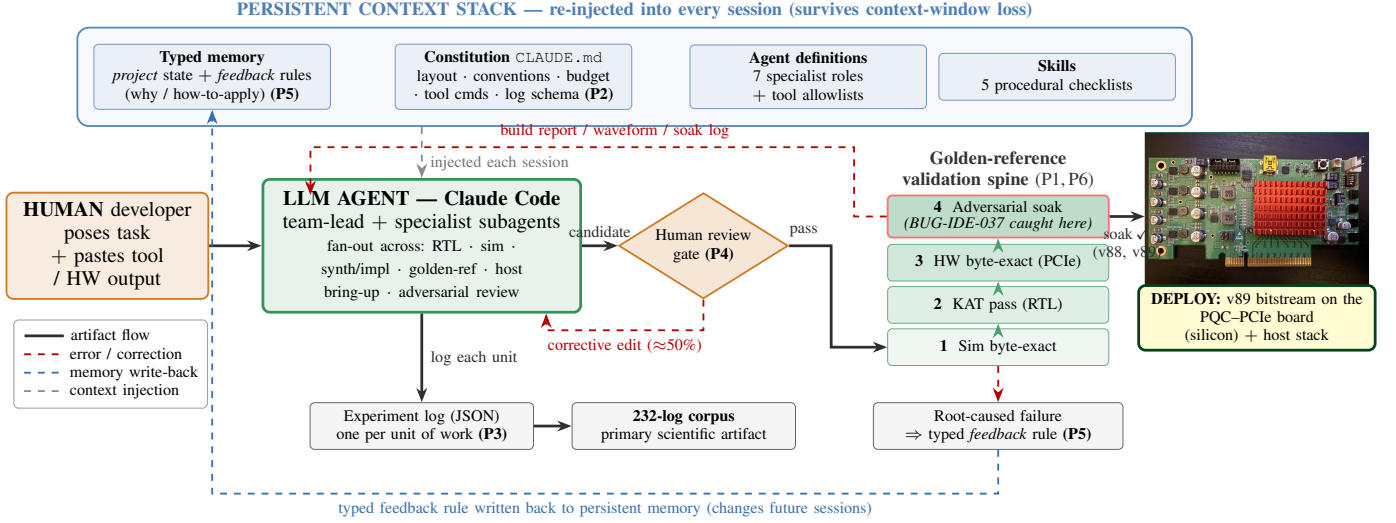
\begin{figure*}[!t]
\centering
\resizebox{\textwidth}{!}{%
\begin{tikzpicture}[
  yscale=0.86,
  >={Stealth[length=2.2mm]},
  ctx/.style={rounded corners=2pt, draw=darkblue!55, fill=darkblue!8, align=center, inner sep=3pt, font=\scriptsize, text width=34mm},
  human/.style={rounded corners=3pt, draw=darkorange!85, thick, fill=darkorange!12, align=center, inner sep=4pt, font=\small, text width=28mm},
  agent/.style={rounded corners=3pt, draw=darkgreen!80, very thick, fill=darkgreen!10, align=center, inner sep=5pt, font=\small, text width=45mm},
  gate/.style={diamond, aspect=1.6, draw=darkorange!85, thick, fill=darkorange!16, align=center, inner sep=1pt, font=\scriptsize, text width=16mm},
  rung/.style={rounded corners=2pt, draw=darkgreen!60, align=center, inner sep=2.5pt, font=\scriptsize, text width=32mm, minimum height=5.6mm},
  deliv/.style={rounded corners=3pt, draw=darkgreen!55!black, very thick, fill=yellow!22, align=center, inner sep=4pt, font=\small, text width=26mm},
  logbox/.style={rounded corners=2pt, draw=black!60, fill=black!4, align=center, inner sep=3pt, font=\scriptsize, text width=32mm},
  flow/.style={->, very thick, black!82},
  err/.style={->, thick, dashed, red!72!black},
  mem/.style={->, thick, dashed, darkblue!78},
  ctxflow/.style={->, thick, dashed, black!45}
]

\node[ctx] (lmem) at (3.1,8.0)  {\textbf{Typed memory}\\\emph{project} state $+$ \emph{feedback} rules (why / how-to-apply)~\textbf{(P5)}};
\node[ctx] (lcon) at (7.7,8.0)  {\textbf{Constitution} \reg{CLAUDE.md}\\layout $\cdot$ conventions $\cdot$ budget $\cdot$ tool cmds $\cdot$ log schema~\textbf{(P2)}};
\node[ctx] (lag)  at (12.2,8.0) {\textbf{Agent definitions}\\7 specialist roles $+$ tool allowlists};
\node[ctx] (lsk)  at (16.0,8.0) {\textbf{Skills}\\5 procedural checklists};
\begin{scope}[on background layer]
  \node[rounded corners=5pt, draw=darkblue!65, thick, fill=darkblue!3,
        fit=(lmem)(lcon)(lag)(lsk), inner sep=7pt] (band) {};
\end{scope}
\node[font=\bfseries\footnotesize, text=darkblue!70, anchor=south] at (band.north)
  {PERSISTENT CONTEXT STACK --- re-injected into every session (survives context-window loss)};

\node[human] (H) at (1.5,5.0)  {\textbf{HUMAN} developer\\poses task\\$+$ pastes tool / HW output};
\node[agent] (A) at (6.3,5.0)  {\textbf{LLM AGENT --- Claude Code}\\team-lead $+$ specialist subagents\\{\scriptsize fan-out across: RTL $\cdot$ sim $\cdot$ synth/impl $\cdot$ golden-ref $\cdot$ host bring-up $\cdot$ adversarial review}};
\node[gate]  (G) at (10.6,5.0) {Human review gate~\textbf{(P4)}};
\node[rung, fill=darkgreen!6]  (rSim)  at (15.1,3.20) {\textbf{1}~~Sim byte-exact};
\node[rung, fill=darkgreen!12] (rKAT)  at (15.1,3.95) {\textbf{2}~~KAT pass (RTL)};
\node[rung, fill=darkgreen!18] (rHW)   at (15.1,4.70) {\textbf{3}~~HW byte-exact (PCIe)};
\node[rung, fill=darkgreen!26, draw=red!55, thick] (rSoak) at (15.1,5.52)
  {\textbf{4}~~Adversarial soak\\[-1pt]{\scriptsize\itshape (BUG-IDE-037 caught here)}};
\draw[->, semithick, darkgreen!70] (rSim.north)  -- (rKAT.south);
\draw[->, semithick, darkgreen!70] (rKAT.north)  -- (rHW.south);
\draw[->, semithick, darkgreen!70] (rHW.north)   -- (rSoak.south);
\node[font=\footnotesize, align=center, text=black!80] at (15.1,6.35)
  {\textbf{Golden-reference}\\\textbf{validation spine}~(P1,\,P6)};
\node[draw=darkgreen!55!black, line width=0.7pt, inner sep=0pt] (D) at (19.05,5.52)
  {\includegraphics[width=3.4cm]{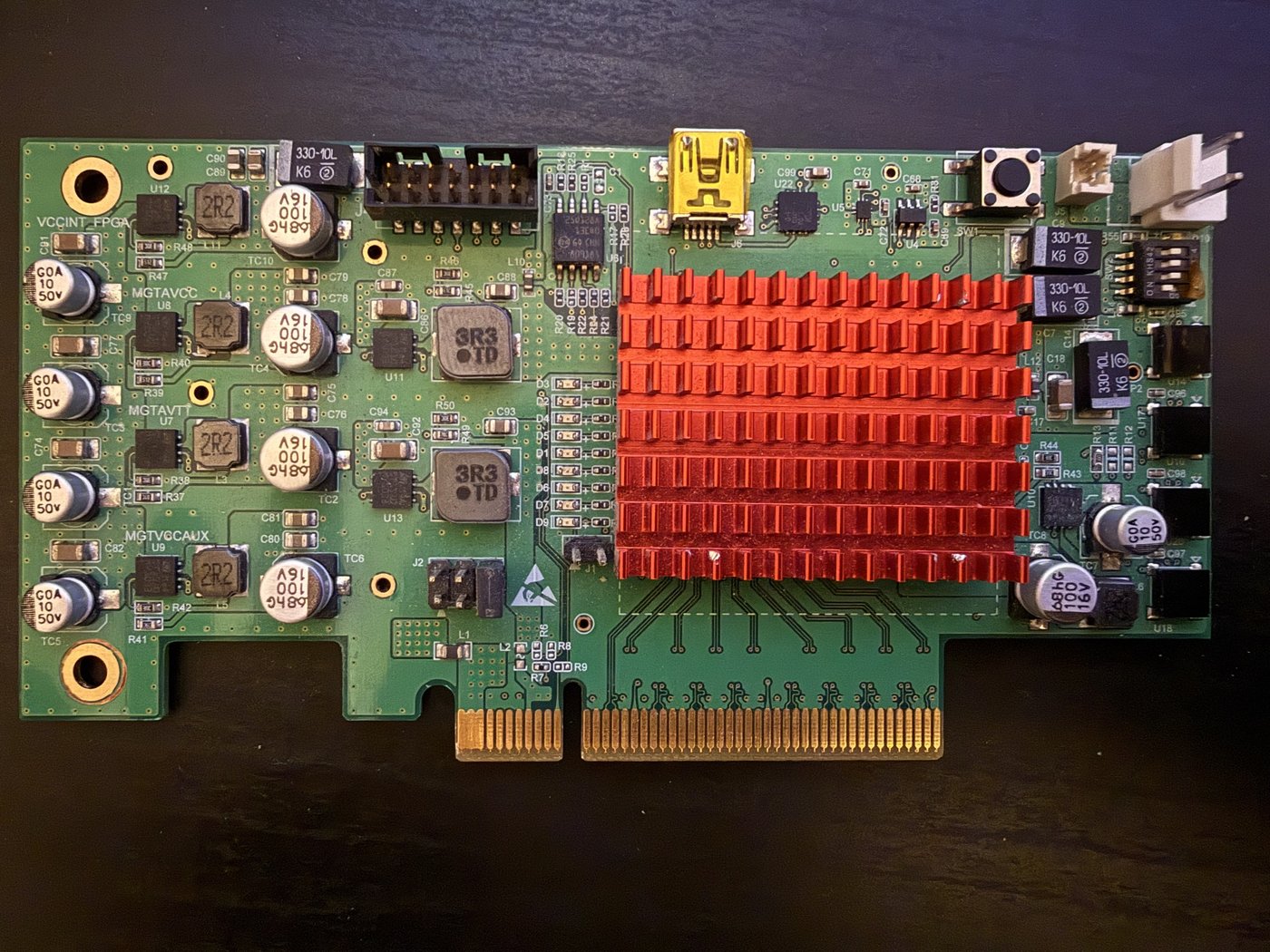}};
\node[deliv, font=\scriptsize, text width=34mm, inner sep=3pt] (Dlbl) at (19.05,3.72)
  {\textbf{DEPLOY:} v89 bitstream on the\\PQC--PCIe board (silicon) $+$ host stack};

\draw[flow] (H.east) -- (A.west);
\draw[flow] (A.east) -- node[above, font=\scriptsize] {candidate} (G.west);
\draw[flow] (G.east) -- node[above, pos=0.35, font=\scriptsize] {pass} (12.55,5.0) |- (rSim.west);
\draw[flow] (rSoak.east) -- node[below=1pt, font=\scriptsize, align=center] {soak\,$\checkmark$\\(v88, v89)} (D.west |- rSoak.east);

\draw[ctxflow] (band.south -| A.north) -- node[right, pos=0.72, font=\scriptsize, text=black!55] {injected each session} (A.north);

\draw[err] (rSoak.west) -- (12.9,5.52) -- (12.9,6.75) -- node[above, pos=0.44, font=\scriptsize] {build report / waveform / soak log} (4.6,6.75) -- (4.6,5.9);
\draw[err] (G.south) -- (10.6,3.5) -- node[below, font=\scriptsize] {corrective edit ($\approx$50\%)} (8.2,3.5) -- (8.2,3.5 |- A.south);

\node[logbox] (LOG) at (6.3,1.8)  {Experiment log (JSON)\\one per unit of work~\textbf{(P3)}};
\node[logbox] (COR) at (10.3,1.8) {\textbf{232-log corpus}\\primary scientific artifact};
\node[logbox, text width=36mm] (FBR) at (15.1,1.8) {Root-caused failure\\$\Rightarrow$ typed \emph{feedback} rule~\textbf{(P5)}};

\draw[flow] (A.south) -- node[right, font=\scriptsize] {log each unit} (LOG.north);
\draw[flow] (LOG.east) -- (COR.west);
\draw[err]  (rSim.south) -- (FBR.north);

\draw[mem] (FBR.south) -- (15.1,0.6) -- node[below, font=\scriptsize] {typed feedback rule written back to persistent memory (changes future sessions)} (3.1,0.6) -- (lmem.south);

\node[draw=black!40, rounded corners=2pt, align=left, font=\scriptsize, inner sep=4pt, fill=white] at (1.55,2.9)
  {\tikz[baseline=-0.6ex]{\draw[flow,-] (0,0) -- (6mm,0);}~artifact flow\\
   \tikz[baseline=-0.6ex]{\draw[err,-] (0,0) -- (6mm,0);}~error / correction\\
   \tikz[baseline=-0.6ex]{\draw[mem,-] (0,0) -- (6mm,0);}~memory write-back\\
   \tikz[baseline=-0.6ex]{\draw[ctxflow,-] (0,0) -- (6mm,0);}~context injection};

\end{tikzpicture}%
}
\caption{The LLM-assisted hardware-design methodology as a closed
\emph{generate--validate--learn} loop ending in deployed silicon.  Tags
\textbf{P1}--\textbf{P6} mark the reusable-protocol steps of
Section~\ref{subsec:protocol}.  The rung-4 pass feeding deployment is
the v88 baseline's 779{,}945-check soak; the shipped v89 re-passed
the full ladder, adversarial soak included
(Section~\ref{sec:threats}, H3).}
\label{fig:loop}
\end{figure*}

Figure~\ref{fig:loop} gives the whole picture.  Candidates passing the human review gate climb the
byte-exact golden-reference validation spine of
Section~\ref{subsec:goldenspine}.

\subsection{The Assistant and Its Toolchain}
\label{subsec:assistant}

The design assistant is Claude Code, an agentic command-line interface
giving the LLM file-system and shell tools and the ability to spawn
specialist subagents.  On wide-scope integration days an Opus-4.6
team-lead orchestrated multiple Sonnet-4.6 subagents.  All models were
Claude-family:
Opus-4.8 (93 logs, 17 with a 1M-context window), Opus-4.7 (63),
Opus-4.6 (47), Opus-4.5 (24), Sonnet-4.6 (3, specialist subagents),
and Fable-5 (1, deep research); counts are per-log model
attributions (one log lacks a tier).

The assistant wrote Verilog-2001 RTL in
\texttt{rtl/}.  It ran Icarus Verilog (iverilog) simulation with
GTKWave review on failures, and Vivado~2024.2 batch synthesis and
implementation via assistant-generated Tcl
(\reg{ExtraNetDelay\_high} placement for congested respins, per-clock
Intra-Clock-Table timing analysis).  It maintained Python golden
references for FIPS~203 and FIPS~204 (round-3 wire formats through
v88, FIPS-final thereafter; Section~\ref{subsec:lineage}),
byte-compared against simulation and silicon.  It drove Python
(\reg{pqc\_pcie\_host.py}) and C99 (\texttt{sw/c/}) host libraries
over XDMA PCIe DMA.  A separate agent thread ran multi-agent
adversarial review of critical design premises, catching at least
one wrong optimization premise (EXP-20260604-001,
Section~\ref{subsec:wrongepisodes}).

\subsection{Context Architecture: The Persistent Markdown Stack}
\label{subsec:context}

Prompts were typically one to three sentences plus pasted tool or
hardware output; the harness injects a four-layer stack of
persistent Markdown files into every session---three layers
versioned in the repository, plus a typed memory in the harness's
per-project store.

\begin{itemize}[nosep]
  \item \textbf{Project constitution (\reg{CLAUDE.md},
    ${\approx}480$ lines; P2).}  Repository layout, Verilog
    conventions, the XC7K160T resource budget, canonical tool
    commands, the logging template of Section~\ref{subsec:logging},
    and both schemes' mathematical parameters.
  \item \textbf{Typed persistent memory (42 topic files at the
    2026-06-26 corpus snapshot, plus a one-line-per-entry index;
    P5).}  Each file
    records a single fact in YAML front matter, written unprompted by
    the assistant.  The 21 \emph{project} files hold
    live status; the 21 \emph{feedback} files hold failure-distilled
    engineering rules with explicit \emph{why} and
    \emph{how-to-apply} fields---e.g., ``on a
    ${\approx}86\%$-LUT, sub-0.1-ns-WNS design, even a no-op
    scaffolding port can break unrelated timing,'' which later prevented repeat failures.
  \item \textbf{Specialist-agent definitions (seven role files).}  Role
    prompts with restricted tool allowlists. The fan-outs of
    Section~\ref{subsec:assistant} instantiate these.
  \item \textbf{Procedural skills (five skill files).}  Checklists
    invoked as named commands---e.g., \reg{/update-all} and
    \reg{/spawn-team}---making the logging discipline of
    Section~\ref{subsec:logging} mechanical.
\end{itemize}

In our experience the \emph{feedback} layer carries the most
methodological value: a lesson learned in week~$n$ changes assistant
behavior in week~$n{+}k$ (Section~\ref{subsec:wrongepisodes}, lesson
themes).  Because the layers outlive any conversation, work resumes
after total context loss without re-teaching.

\begin{figure}[t]
  \centering
\begin{tikzpicture}[x=0.88cm,y=0.90cm,
  every node/.style={inner sep=0pt,outer sep=0pt},
  font=\scriptsize]

\fill[bandc] (4,0.12) rectangle (5,2.44);

\foreach \gx in {0,1,2,3,4,5,6}{
  \draw[gridc,line width=0.3pt] (\gx,0.12) -- (\gx,2.44);
}
\draw[edgec,line width=0.5pt] (0,0.06) -- (6,0.06);

\newcommand{\phasebar}[6]{%
  \fill[#4] (#2,#1-0.16) rectangle (#3,#1+0.16);
  \ifnum#5=1
    \fill[pattern={Lines[angle=45,distance=2.4pt,line width=0.3pt]},pattern color=edgec]
      (#2,#1-0.16) rectangle (#3,#1+0.16);
  \fi
  \draw[edgec,line width=0.4pt] (#2,#1-0.16) rectangle (#3,#1+0.16);
  \node[anchor=east,txtc] at (-0.16,#1) {#6};
}
\phasebar{2.28}{0}{1}{p1c}{0}{\,1\ Foundation}
\phasebar{1.88}{1}{3}{p2c}{0}{\,2\ Core Modules}
\phasebar{1.48}{2}{3}{p3c}{0}{\,3\ Integration}
\phasebar{1.08}{3}{4}{p4c}{0}{\,4\ Validation}
\phasebar{0.68}{4}{5}{p5c}{1}{\,5\ PCIe Bring-up}
\phasebar{0.30}{4}{6}{p6c}{1}{\,6\ PQC-HSM}

\newcommand{\monthlab}[3]{%
  \node[anchor=north,txtc] at (#1,0.00){#2};
  \node[anchor=north,dimc,font=\tiny] at (#1,-0.22){#3};
}
\monthlab{0.5}{Jan}{100}
\monthlab{1.5}{Feb}{92}
\monthlab{2.5}{Mar}{82}
\monthlab{3.5}{Apr}{88}
\monthlab{4.5}{May}{41}
\monthlab{5.5}{Jun}{65}
\node[anchor=east,dimc,font=\tiny] at (-0.16,-0.30){task-pass \%};
\node[anchor=west,dimc,font=\tiny] at (6.10,0.00){2026};

\draw[edgec,line width=0.35pt] (0.9,2.64)--(6.0,2.64);
\newcommand{\dmil}[3]{%
  \node[diamond,draw=edgec,fill=white,line width=0.4pt,
        minimum width=4pt,minimum height=5.5pt] at (#1,2.64){};
  \ifnum#2=1 \node[anchor=south,txtc,font=\tiny] at (#1,2.71){#3};\fi
  \ifnum#2=2 \node[anchor=south,txtc,font=\tiny] at (#1,2.89){#3};\fi
}
\dmil{1.05}{2}{M1}
\dmil{1.50}{1}{M2}
\dmil{2.05}{2}{M3}
\dmil{2.38}{1}{M4}
\dmil{2.62}{2}{M5}
\dmil{2.90}{1}{M6}
\dmil{3.40}{2}{M7}
\dmil{3.78}{1}{M8}

\node[starc] at (4.12,2.64){$\bigstar$};
\node[anchor=south,starc,font=\tiny] at (4.12,2.88){PCIe};
\node[starc] at (6.00,2.64){$\bigstar$};
\node[anchor=south east,starc,font=\tiny] at (6.06,2.88){v89/v1.1 (07-08)};


\end{tikzpicture}
  \caption{Campaign timeline of the deployed Kintex-7~160T work
  (2026-01--06): six phase bars, planned milestones M1--M8 (diamonds),
  and two deployment landmarks (stars; the v89/v1.1 release of
  2026-07-08 is drawn at the axis edge), all detailed in
  \S\ref{subsec:context}.  The faint monthly task-pass~\%
  (\S\ref{subsec:overtime}) reflects task mix, not a learning trend
  ($n{=}1$).}
  \label{fig:timeline}
\end{figure}
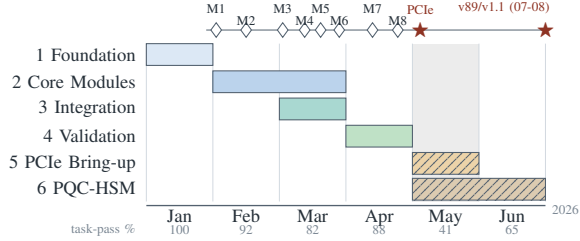

One further persistent artifact sat outside the stack: a
plan-and-milestone roadmap in the project dashboard's timeline data,
decomposing the six-month effort into an ordered sequence (M1--M8bp,
76~entries; Fig.~\ref{fig:timeline}).
Each milestone's ``done'' is verifiable, so a resuming session
inherits the last closed increment and the next target.  M1--M8 were
fixed through the v1.0 release; M8a--M8bp were appended afterward for
the unanticipated silicon bring-up, key-custody, and productization
work.  We therefore read its role as purely
\emph{structural} and do \emph{not} claim the
plan raised the per-task success rate
($n{=}1$; Section~\ref{subsec:overtime}).

In Figure~\ref{fig:timeline}, M1--M8 span
pre-silicon to first release (weeks~3--12): architecture locked
(M1), NTT verified (M2), all modules done (M3), Kyber (M4) and
Dilithium (M5) integrated (campaign-era scheme names), optimization
(M6) and validation (M7) complete, and the accelerator-only
repository release (M8; distinct from the productized Release~v1.0 of
Section~\ref{subsec:lineage}).  M8a--M8bp
group into five themes: (i)~\emph{PCIe hardware bring-up}: a
Gen2$\times$8 link and live XDMA (M8a; first bitstream 2026-05-04),
then all six PQC operations byte-exact at full back-to-back PCIe
throughput.  (ii)~\emph{On-chip entropy}: an HRNG audit, QRNG/hybrid
modes, and an internal-seed KeyGen whose seed never crosses PCIe.
(iii)~\emph{Cycle reduction}: the 125~MHz v69 respin and the v70--v77
datapath work.  (iv)~\emph{On-die key custody} (v79--v87):
sign-from-vault, on-chip key-wrap/export, a KEK rooted in the FPGA's
factory-programmed Device-DNA identifier, and SHAKE-KDF, capped by a TLS-1.3 CertificateVerify through hardware
custody.  (v)~\emph{Productization and hardening}: the BUG-IDE-037 fix
in v84, exhaustive and eight-hour soaks, a native-C host library,
the v88 Am-241 quantum-RNG baseline, and the v1.0 product
release.  The campaign closed with the FIPS-203/204 finalization---firmware v89 and the
v1.1 release (ML-KEM-768/ML-DSA-65)---Fig.~\ref{fig:timeline}'s second
deployment star.

\subsection{Experiment-Logging Discipline}
\label{subsec:logging}

Every significant unit of work was recorded as structured JSON in
\texttt{llm\_logs/experiments/}, following the nine-field schema of
Section~\ref{subsec:template}; per-field presence accompanies each
count in Section~\ref{sec:empirical}.  Logging was enforced at
experiment close, not retrospectively.

\textbf{Data availability.}  The experiment logs (JSON), milestone
index, 40-category-to-ten-bucket mapping, and the scripts producing
the tables of Section~\ref{sec:empirical} are available from the
authors, credential-redacted; the logging schema, the methodology template of
Section~\ref{subsec:template}, and the complete v89
hardware-validation transcript (Section~\ref{subsec:lineage}) are
supplementary material.

\subsection{The Golden-Reference Validation Spine}
\label{subsec:goldenspine}

No RTL reached a hardware build without byte-exact agreement with the
Python golden reference in simulation.  The promotion ladder was: Sim byte-exact $\to$ KAT pass (RTL)
$\to$ HW byte-exact (PCIe) $\to$ adversarial soak.
The adversarial soak (Section~\ref{subsec:bugide037}, reliability
closure) exercises randomized message classes outside the KAT
vectors; it caught BUG-IDE-037, invisible to all earlier rungs.

\subsection{Human-in-the-Loop Role}
\label{subsec:hitl}

The developer (the first author) posed tasks, \emph{gated} every
generated artifact (no output entered the simulator, synthesis, or the
board without review; ${\approx}50\%$ of edit-tracked generations
needed a corrective edit, Section~\ref{subsec:hitldata}),
supplied build reports and
waveforms as corrective feedback, decided when to roll back, and
made all \emph{real hardware decisions}.  The claim under test
is that the agentic loop \emph{with these human gates}---not
unaided model authorship---carried the complete flow to deployment.

\subsection{A Reusable Protocol for Other Implementations}
\label{subsec:protocol}

Applying the loop to another scheme, accelerator domain, or
FPGA/ASIC flow takes six steps:

\begin{enumerate}[nosep, label=\textbf{P\arabic*.}, leftmargin=*]
  \item \textbf{Executable golden reference first.}  Before any RTL,
    byte-exact agreement with the reference (here, Python
    FIPS~203/204) is the sole acceptance criterion for every artifact
    (Section~\ref{subsec:goldenspine}).
  \item \textbf{Write the constitution.}  One versioned Markdown file
    (Section~\ref{subsec:context}) of layout, conventions, budget,
    tool commands, and logging schema.
  \item \textbf{Instrument before generating.}  Adopt the experiment
    log (Section~\ref{subsec:logging}) from day one.
  \item \textbf{Gate by hardware coupling.}  Every artifact crosses
    the review gate (Fig.~\ref{fig:loop}); concentrate review
    \emph{depth} where physical coupling is high (the 50--53\%
    buckets of Table~\ref{tab:successbucket}), lighter-touch where low
    (77--85\%).  Here every artifact was gated
    (Section~\ref{subsec:hitl}); the coupling--success ordering is
    offered as a default prior, not a validated predictor.
  \item \textbf{Convert every root-caused failure into a typed feedback
    rule} with why/how-to-apply fields
    (Section~\ref{subsec:context}).
  \item \textbf{Promote only up the ladder} of
    Section~\ref{subsec:goldenspine}; never skip a rung.
\end{enumerate}

The scheme-specific investment was the golden reference and KAT
vectors.  The device-specific investment was the constitution's
tool-command and resource-budget sections.  Everything else (the
memory-stack \emph{schema}, agent roles, logging schema, and
promotion ladder) remained unchanged through the corpus-frozen v88
and the shipped v89.
Whether it transfers with a similar success profile is untested
(Section~\ref{sec:threats}, H1--H2).

\subsection{The Methodology as a Reusable Template}
\label{subsec:template}

The concrete artifact of the protocol in
Section~\ref{subsec:protocol} is a repository-skeleton template
holding the constitution \texttt{CLAUDE.md} (P2), the golden
reference \texttt{ref/} (P1), the typed-memory files (P5), the
agent-role and skill \emph{scaffolds}, and the JSON experiment-log
schema (P3).  Only \texttt{CLAUDE.md} and \texttt{ref/} are
rewritten per target; the remaining layers copy unchanged, and the
memory contents and concrete role/skill files accrue as the project
earns them (the template ships one example of each).

\section{Empirical Results}
\label{sec:empirical}

\subsection{Corpus and Overall Success Rate}
\label{subsec:corpus}

The 232-experiment corpus, frozen at the 2026-06-26 analysis
snapshot, records one intensive campaign (2026-01 to 2026-06)
spanning 40 task categories and 76 named milestones
(M1--M8bp).\footnote{Eight planned milestones (M1--M8, through the v1.0
release) plus 68 finer-grained post-plan sub-milestones (M8a--M8bp, in
spreadsheet-column labeling) tracking the silicon bring-up, key-custody,
and productization work.}  Of the 232, 166 score success and 66
failure, an overall success rate of \textbf{71.6\%} (Wilson 95\%
interval $65$--$77\%$).  Generated or
modified code sums to ${\approx}26{,}373$ lines.

\texttt{syntax\_correct True} appears in 166 logs,
\texttt{simulation\_pass True} in 107, \texttt{synthesis\_pass True}
in 106, and \texttt{functional\_correctness passed} in 118 of the 127
logs carrying the field.  Text explicitly claiming ``HW-validated,''
``byte-exact,'' or ``silicon'' appears in 93 logs.

An experiment scores \emph{success} when its log records the task's
acceptance criterion met at close, \emph{after} any in-experiment
iteration and human edits: functional correctness where a functional
oracle exists (byte-exact against the FIPS reference, a passing
simulation, or on-silicon validation), the applicable pass field
otherwise.  Abandoned or rolled-back outcomes score failure.  The
49.2\% manual-edit rate of Section~\ref{subsec:hitldata} measures
intervention \emph{en route}.
The \texttt{syntax\_correct} field is not an independent gate: it is
recorded (always \texttt{True}) exactly in the
166 accepted logs and omitted in the 66 failures, so its count
coinciding with the success count is definitional.

\subsection{Success by Task Category}
\label{subsec:bycategory}

Table~\ref{tab:successbucket} presents the ten task buckets.

\begin{table}[t]
\caption{Success rate by task category (232 experiments).  \emph{Model} is
         the most-used Claude tier(s) logged per category.}
\label{tab:successbucket}
\centering
\small
\setlength{\tabcolsep}{4pt}
\begin{tabular}{lrrrc}
\toprule
Category & $N$ & Success & Rate & Model \\
\midrule
Documentation               & 53 & 45 & 85\% & 4.5--4.8 \\
Host-software / tooling     & 10 &  8 & 80\% & 4.7/4.8 \\
Release / infra / PM        & 14 & 11 & 79\% & 4.6/4.8 \\
Research / architecture     & 13 & 10 & 77\% & 4.5/4.6 \\
Bug-fix / debug             & 47 & 35 & 74\% & 4.7 \\
RTL generation              &  7 &  5 & 71\% & 4.7 \\
Verification                & 33 & 23 & 70\% & 4.8 \\
Review / audit              &  3 &  2 & 67\% & 4.8 \\
\textbf{Synthesis / optimization} & \textbf{38} & \textbf{20} & \textbf{53\%} & \textbf{4.8} \\
\textbf{Integration / bring-up}   & \textbf{14} &  \textbf{7} & \textbf{50\%} & \textbf{4.8} \\
\midrule
\textbf{Total}              & \textbf{232} & \textbf{166} & \textbf{71.6\%} & --- \\
\bottomrule
\end{tabular}
\end{table}

\textbf{Finding.}  \emph{LLM success rate falls the more directly a
task touches FPGA placement, timing, and live-hardware state.}  We
ranked the ten buckets by coupling to live silicon---documentation
(rank~1, no tool execution) through research/architecture,
release/infra, host-software, review/audit, verification, RTL
generation, bug-fix/debug, and synthesis/optimization to
integration/bring-up (rank~10, live-silicon state required)---a
ranking fixed from the coupling criterion alone (so not circular),
giving Spearman $\rho \approx -0.83$ against the measured rates.  This is a \emph{descriptive} pattern, not a
controlled measurement: the ranking is subjective, the buckets are
neither independent samples nor equal-sized, and the corpus is
$n{=}1$, so we attach no significance test.  Model choice is an
unlikely explanation, since the two lowest-success buckets were the
most Opus-4.8-concentrated---our strongest tier
(Table~\ref{tab:successbucket}, \emph{Model} column).  Even so, at
$n{=}1$ coupling and model quality cannot be fully separated
(Section~\ref{sec:threats}, H2).

The ten buckets coarse-grain the 40 logged \texttt{task.category}
labels by keyword (segmentation and \emph{oracle-stringency} caveats:
Section~\ref{sec:threats}, H1).  Per-bucket Wilson 95\% intervals are
wide at these sample sizes (documentation $73$--$92\%$,
synthesis/optimization $37$--$68\%$); only the
documentation--synthesis/optimization intervals are disjoint.

\subsection{Iterations and One-Shot Rate}
\label{subsec:iterations}

Iteration counts were recorded in 179 of 232 logs.\footnote{An
iteration is one generate--review--correct cycle, from a log's
optional count field (\reg{iterations\_needed}, or
\reg{total\_iterations} in early-schema logs; $1$~=~a one-shot
acceptance).  53 of the 232 experiments in
Table~\ref{tab:successbucket} leave it unrecorded, across many
task categories rather than one type.}  The one-shot rate is 57.0\% (102/179); 74.3\% closed within
two iterations and 93.3\% within five, with mean 2.12, median 1, and
maximum 14.  A 95\% interval on 102/179 spans roughly 50--64\% (53-log
missingness caveat: Section~\ref{sec:threats}, H1).  A higher count reflects rework, not
failure (Section~\ref{subsec:corpus}).

The three hardest tasks by iteration count were EXP-20260324-001 (14
iterations, Dilithium signature top-level integration)\footnote{Log-derived
task names retain the campaign-era CRYSTALS names (Kyber, Dilithium); the
text otherwise uses ML-KEM/ML-DSA for the standardized schemes.}, EXP-20260522-001
(9, PCIe Dilithium Sign+Verify debug), and EXP-20260318-002
(8, Kyber KEM integration, six bugs in
\texttt{kyber\_decaps.v}).

\subsection{Human-in-the-Loop Rate and Footprint}
\label{subsec:hitldata}

\textbf{Correction rate.}  Manual-edit records appear in the \emph{same} 179 logs
that record iteration counts (Section~\ref{subsec:iterations}).
Of the 179, 91 (50.8\%)
report ``None'' (accepted as produced)
and 88 (49.2\%) at least one human correction.  Intervention types
(among the 87 with fine-grained tags; multi-tag, sums exceed 87):
logic/RTL race correction 24,
documentation fix 19, host-SW fix 18, syntax/compile fix 17,
parameter tuning 15, timing/placement adjustment 12,
rollback/revert 8, integration wiring 6.

\textbf{Interpretation.}  The largest classes---RTL logic/race
correction, host-SW fixes, timing/placement adjustments---are those
where a software-trained model is structurally weakest, a weighting
directionally consistent with the bucket-level finding
(Table~\ref{tab:successbucket}); corrections nonetheless span all
task types (raw counts; base-rate/task-mix caveat: H2).

\textbf{The human-in-the-loop footprint.}  The 49.2\% edit rate
quantifies only one of five operator roles, which the logs measure
unevenly:
\begin{itemize}[nosep, leftmargin=*]
  \item \emph{Corrective edits} (measurable): the 49.2\% edit rate
    above (worst-case missing-not-at-random bounds $37.9$--$60.8\%$ over
    the full 232, from the 53 unrecorded logs) and the 2.12 mean
    iterations of Section~\ref{subsec:iterations}.
  \item \emph{Rollback / restart decisions} (partly measurable): 8 logged
    reverts.
  \item \emph{Go/no-go hardware acts} (weakly measurable): every flashed
    bitstream and accepted timing closure is an irreversible operator
    decision---dozens across the bitstream lineage through v88---but not
    individually logged.
  \item \emph{Task specification} and \emph{artifact review/gating} (not
    measurable as a fraction): the operator wrote every prompt, gated
    every artifact entering the toolchain (Section~\ref{subsec:hitl}),
    and committed 100\% of changes---uniform across \emph{all}
    experiments, and plausibly the largest human inputs.
\end{itemize}
Self-recording and authorship-apportionment caveats are stated as
threats (Section~\ref{sec:threats}, H1--H2).

\subsection{Bug Taxonomy}
\label{subsec:bugtaxonomy}

Table~\ref{tab:bugtax} enumerates the project's eight defect
classes.  Each validation layer catches a class the others miss, and
the blindness is instrument-specific: simulation misses what only
silicon shows, KAT misses what only randomization shows, and
synthesis silently mis-elaborates RTL that simulates clean.  Three of
these families dominate the bring-up record and are analyzed in
Section~\ref{subsec:bringup}.

\begin{table*}[t]
\caption{Bug taxonomy of defect classes encountered across the campaign,
         with detection method.}
\label{tab:bugtax}
\centering
\begin{tabular}{@{}p{4.2cm}p{6.3cm}p{5.9cm}@{}}
\toprule
Class & Representative instances & Detection method \\
\midrule
Non-blocking-assignment (NBA) / registered-vs-combinational handshake race &
  Multi-write BRAM (Vivado drops one write; BUG-IDE-026); shift-by-1
  corruption from combinational \reg{req\_out} &
  Simulation (iverilog); Vivado silently mis-synthesizes iverilog-clean RTL \\[2pt]
Host-vs-RTL misdiagnosis &
  BUG-IDE-031/033/034: apparent RTL regressions were host-side
  (Section~\ref{subsec:bringup}).  BUG-IDE-036: retained Verify-FSM
  state across back-to-back calls---RTL-behavioral (reproduced in pure
  RTL simulation, 2026-08), mitigated by the per-call soft-reset (D1) &
  Live hardware A/B testing \\[2pt]
Timing-closure / placement congestion &
  XDMA \reg{userclk1} 500\,MHz path tipped negative; fixed by Explore
  seed-sweep, not \reg{phys\_opt} &
  Build timing report (per-clock Intra-Clock-Table) \\[2pt]
FIPS / canonical-encoding mismatch &
  BUG-IDE-035: \texttt{tr} = SHAKE256 of the 1952-B FPGA pk vs.\ the
  3104-B Python packed pk &
  Hardware byte-diff against reference \\[2pt]
Data-dependent crypto logic (norm-check off-by-N) &
  BUG-IDE-037: norm check read too few coefficients under BRAM latency;
  rare out-of-bound signature &
  8-hour adversarial soak; invisible to KAT vectors \\[2pt]
Multi-block host truncation &
  On-chip keygen-wrap: \texttt{s2[1-5]} wrong; host returned on pk TLAST before
  on-chip recovery finished &
  Hardware byte-diff (last blocks wrong, rest exact) \\[2pt]
Cross-process / driver transport &
  XDMA card-to-host (C2H) engine backlog ($\approx$1041\,B) surviving
  close$\to$open; kernel-5.15 zero-padding partial C2H reads &
  Hardware A/B across hosts/kernels \\[2pt]
Documentation over-claim &
  Stale ``5--18$\times$ cycle gap'' (unconfirmed comparators); stale
  ``125\,MHz'' clock label in five deployed claims &
  Post-edit grep audit; review \\
\bottomrule
\end{tabular}
\end{table*}

\subsection{Ten Episodes of ``Confidently Wrong, Then Corrected''}
\label{subsec:wrongepisodes}

In ten episodes the LLM (or the developer acting on its output)
reached a confident, incorrect conclusion, acted on it, and later
corrected course---the hazard of plausible but wrong diagnosis.  Five in detail:

(1)~\textbf{Premature ``zero-failures'' claim} (EXP-20260614-001;
corrected in EXP-20260615-001): a
${\approx}84{,}000$-check soak prompted it; later soaks surfaced
BUG-IDE-037, misdiagnosed twice until proven deterministic
(Section~\ref{subsec:bugide037}).
(2)~\textbf{A ``driver flake'' that was hardware} (EXP-20260619-001):
BUG-IDE-021, managed for months as such, was a stale \reg{HSM\_CMD}
level register surviving a soft-reset.
(3)~\textbf{Draining the wrong FIFO} (EXP-20260620-002): a stuck C2H (card-to-host)
path drew two wrong ``drain the FIFO'' remedies; $\approx$1041\,B sat in
the C2H \emph{engine}, not the char FIFO---fixed by a burn-op
(\reg{\_warmup\_c2h}).
(4)~\textbf{``LUT-bound'' that was placement density}
(EXP-20260604-001): a multi-agent adversarial review overturned the
``LUT-bound / idle DSPs'' premise; the actual constraint was 97.49\%
slice occupancy.
(5)~\textbf{An ``occasional'' bug that was deterministic}
(EXP-20260529-001): Dilithium Verify failures called ``occasional
contamination'' were 200/200 deterministically wrong (BUG-IDE-036,
Table~\ref{tab:bugtax}).
The other five: EXP-20260625-001 (``other board broken''---it was the
best); EXP-20260619-004 (a \reg{ram\_style="block"} fix staged, then
infeasible on fallback); EXP-20260617-002 (a wrong on-die \texttt{sk}
blamed on RTL; two sims showed host truncation); EXP-20260613-002 (a KDF
``would not fit'' the K160T; empirically wrong); EXP-20260422-006 (a
correct BRAM fix reverted, then un-reverted).

\textbf{Lesson themes.}  Manual thematic coding (not a mechanical
keyword count) of \texttt{lessons\_learned} across the 232 logs yields eight themes (approximate log tallies in parentheses;
multi-theme, so tallies sum past 232), each stated as an applicable rule:
(i)~\emph{verify the specific buggy phrase or count, not the bare
token} (75)---generic searches always ``pass''; grep the exact wrong
string and check its count;
(ii)~\emph{registered-vs-combinational / NBA discipline} (73)---a
handshake output derived from registered state must itself be
registered; iverilog tolerates what silicon will not
(Section~\ref{subsec:bugtaxonomy});
(iii)~\emph{timing-as-congestion} (71)---near the area ceiling,
negative slack is placement luck: sweep Explore seeds from one
checkpoint, not \reg{phys\_opt}; read the per-clock timing tables;
(iv)~\emph{host-vs-RTL boundary discipline} (37)---a boundary symptom
does not name the layer that produced it; A/B the host-side variable
before any RTL hunt (Section~\ref{subsec:bringup});
(v)~\emph{read the build report, not historical docs} (35)---area and
clock numbers go stale in prose, never in the report;
(vi)~\emph{honest scoping} (29)---state the validation rung that
earns a claim and claim nothing above it;
(vii)~\emph{adversarial review to challenge premises} (28)---the
costliest errors were confident premises, so red-team the diagnosis,
not just the patch; and
(viii)~\emph{a golden-reference sim before hardware} (22)---a
byte-exact, X-clean model of every change before any bitstream.

\subsection{Success Rate Over Time}
\label{subsec:overtime}

Monthly success (Fig.~\ref{fig:timeline}) ran 100\% (22/22), 92\%
(12/13), 82\% (14/17), and 88\% (45/51) over 2026-01 through 2026-04.
It then dipped sharply to 41\% (18/44) in 2026-05, the hardest phase,
which combined PCIe Dilithium Sign$+$Verify bring-up, the HRNG audit,
and a cycle-optimization campaign with several sub-floor or rolled-back
builds.  It recovered to 65\% (55/85) in 2026-06 with the shift back to
hardening and documentation.

\textbf{Interpretation.}  The trajectory is not a learning curve.  Live-hardware
tasks were impossible before the first PCIe bitstream (2026-05-04, Section~\ref{subsec:bringup}), so the
monthly trajectory is largely the coupling gradient of
Table~\ref{tab:successbucket} expressed in time, not an independent
confirmation of it.  Concurrent model-version changes
(Section~\ref{subsec:assistant}) and phase effects such as
accumulating design debt are further confounds $n{=}1$ cannot
separate, though the bucket-level pattern favors the task-mix reading.
This is the empirical form of contribution~\ref{contrib:empirical}.

\subsection{Compute Cost: Token Consumption}
\label{subsec:tokens}

\textbf{Measurement window.}  The harness records per-message token
usage, but transcripts of inactive sessions are pruned after roughly a
month.  What survived was one continuously active session spanning
the final six weeks (2026-05-17 to 2026-07-02): 117 of the 238
experiments logged by the measurement date (six beyond the
frozen 232-experiment snapshot used elsewhere), i.e.\ 49\% of all
logged campaign activity, covering the PCIe custody, soak-hardening,
and documentation phases.

\textbf{Token totals.}  Deduplicating usage records by message ID
across the main session and all 615 subagent and workflow threads gives
17,877 assistant turns (9,303 main-session, 8,574 subagent) processing
${\approx}4.47$ billion tokens: 6.3\,M fresh (uncached) input, 21.6\,M
generated output, 145.1\,M prompt-cache writes, and 4{,}292.2\,M
prompt-cache reads.

\textbf{Cache-dominated.}  99.4\% of processed tokens are prompt-cache
reads and writes.  The
campaign's ${\approx}26$\,kLOC corresponds, at a conservative
${\sim}15$ tokens per line, to under 2\% of the generated tokens in
this window alone (and much of that code predates the window).
Scaling linearly by logged-activity share gives a campaign-order
estimate of ${\sim}9$\,B tokens processed and ${\sim}44$\,M
generated.  The per-experiment token-estimate field is unreliable; the harness
transcripts are ground truth.

\subsection{Measured Productivity, Efficiency, and Accuracy}
\label{subsec:advantages}

The numbers above map onto the three axes that matter to a team
adopting the protocol of Section~\ref{subsec:protocol} (scope: H1):
\begin{itemize}[nosep, leftmargin=*]
  \item \emph{Productivity}---one developer plus the assistant,
    ${\approx}$5--6 months, ${\approx}26{,}373$ LOC generated or
    modified, 232 logged experiments, a deployed accelerator $+$ HSM.
  \item \emph{Efficiency}---57.0\% one-shot acceptance, mean 2.12
    iterations, 21.6\,M generated tokens at 99.4\% cache-mediated
    processing in the measured window (campaign-order ${\sim}44$\,M, an
    extrapolation from the 49\% window).
  \item \emph{Accuracy}---71.6\% overall experiment success (77--85\%
    documentation/research vs.\ 50--53\% synthesis and bring-up), 6/6
    operations byte-exact on silicon, and a 779{,}945-check adversarial
    soak with zero failures.
\end{itemize}

The productivity claim is the artifact's existence and validation
status (Section~\ref{sec:threats}, H1).  On efficiency, the assistant \emph{reads}
two orders of magnitude more than it \emph{writes}
(${\approx}207{:}1$), so the compute cost is context, not generation.  On accuracy, raw acceptance rates are the
wrong lens; correctness comes from the validation spine, and the
coupling gradient motivates the protocol's human gating
(Section~\ref{subsec:protocol}).

\section{The Artifact as Evidence}
\label{sec:artifact}

The 232-experiment methodology produced a silicon artifact that
scales past single modules to a complete deployed cryptosystem.  Three bitstream labels recur: v87
(earlier deployed build, resource baseline only), v88 (the frozen-corpus
deployed baseline behind every Section~\ref{sec:empirical} hardware
statistic),
and v89 (the shipped release with FIPS-final wire formats,
Section~\ref{subsec:lineage},
Tables~\ref{tab:resources}--\ref{tab:comparison}).  A full architectural
exploration and side-channel analysis are out of scope
(Section~\ref{sec:conclusion}); Section~\ref{sec:pcie} adds the PCIe
board, host--device protocol, and productized bitstream lineage.

\subsection{Unified Architecture}
\label{subsec:architecture}

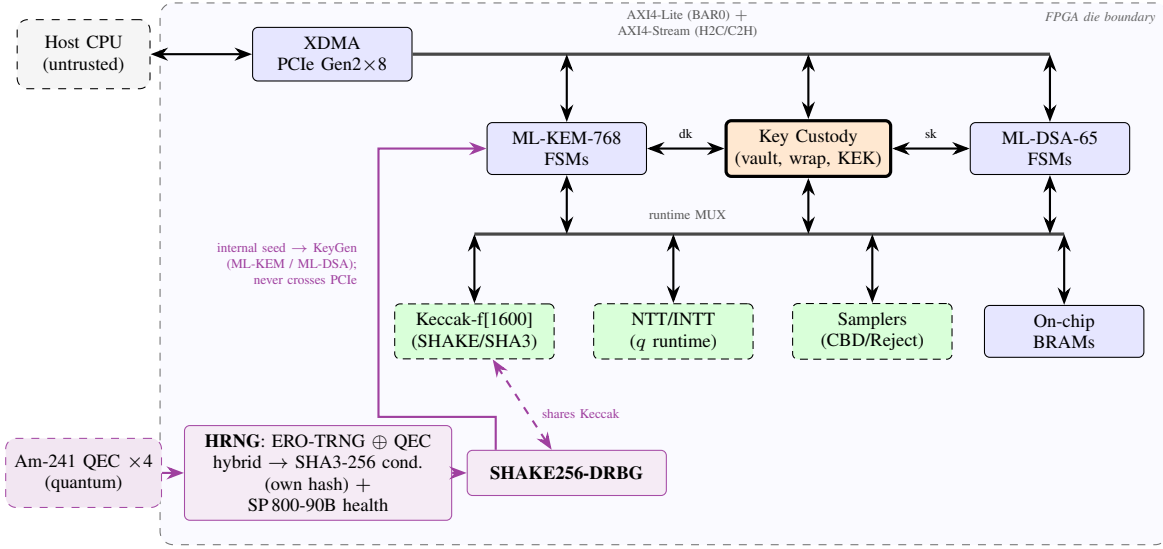
\begin{figure*}[t]
\centering
\begin{tikzpicture}[
  scale=0.94, yscale=0.83, every node/.style={transform shape=false},
  font=\scriptsize,
  block/.style={rectangle, draw, rounded corners=2pt, align=center,
                minimum height=1.7em, minimum width=6.0em, fill=blue!10},
  shared/.style={rectangle, draw, densely dashed, rounded corners=2pt, align=center,
                 minimum height=1.7em, minimum width=6.0em, fill=green!15},
  host/.style={rectangle, draw, dashed, rounded corners=4pt, align=center,
               minimum height=2.6em, minimum width=5.0em, fill=gray!8},
  qsrc/.style={rectangle, draw=violet!70, dashed, rounded corners=4pt, align=center,
               minimum height=2.6em, minimum width=5.0em, fill=violet!10},
  ent/.style={rectangle, draw=violet!70, rounded corners=2pt, align=center,
              minimum height=1.7em, fill=violet!8},
  darr/.style={Stealth-Stealth, thick},
  seed/.style={-Stealth, thick, violet!75},
  bus/.style={line width=1.2pt, black!72},
  >=Stealth
]
\node[host]  (hostbox) at (-5.2,3.6) {Host CPU\\(untrusted)};
\node[block] (xdma)    at (-1.7,3.6) {XDMA\\PCIe Gen2$\times$8};
\draw[darr] (hostbox) -- (xdma);

\node[block] (kyber) at (1.6,2.0) {ML-KEM-768\\FSMs};
\node[block,fill=orange!18,very thick] (hsm) at (5.0,2.0) {Key Custody\\(vault, wrap, KEK)};
\node[block] (dil)   at (8.4,2.0) {ML-DSA-65\\FSMs};
\draw[darr] (kyber) -- node[midway,above]{\tiny dk} (hsm);
\draw[darr] (hsm)   -- node[midway,above]{\tiny sk} (dil);

\draw[bus] (xdma.east) -- (8.4,3.6);
\node[font=\tiny, text=black!65, align=center, anchor=south] at (3.3,3.72)
  {AXI4-Lite (BAR0) $+$\\AXI4-Stream (H2C/C2H)};
\draw[darr] (1.6,3.6) -- (kyber.north);
\draw[darr] (5.0,3.6) -- (hsm.north);
\draw[darr] (8.4,3.6) -- (dil.north);

\draw[bus] (0.3,0.55) -- (8.6,0.55);
\node[font=\tiny, text=black!65, anchor=south] at (3.3,0.64)
  {runtime MUX};
\draw[darr] (kyber.south) -- (1.6,0.55);
\draw[darr] (hsm.south)   -- (5.0,0.55);
\draw[darr] (dil.south)   -- (8.4,0.55);

\node[shared] (keccak) at (0.3,-1.1) {Keccak-f[1600]\\(SHAKE/SHA3)};
\node[shared] (ntt)    at (3.1,-1.1) {NTT/INTT\\($q$ runtime)};
\node[shared] (samp)   at (5.9,-1.1) {Samplers\\(CBD/Reject)};
\node[block]  (mem)    at (8.6,-1.1) {On-chip\\BRAMs};
\draw[darr] (keccak.north) -- (0.3,0.55);
\draw[darr] (ntt.north)    -- (3.1,0.55);
\draw[darr] (samp.north)   -- (5.9,0.55);
\draw[darr] (mem.north)    -- (8.6,0.55);

\node[qsrc] (qec) at (-5.2,-3.5) {Am-241 QEC $\times$4\\(quantum)};
\node[ent, text width=33mm] (hrng) at (-1.9,-3.5)
  {\textbf{HRNG}: ERO-TRNG $\oplus$ QEC\\hybrid $\to$ SHA3-256 cond.\\(own hash) $+$ SP\,800-90B health};
\node[ent, text width=24mm] (drbg) at (1.6,-3.5)
  {\textbf{SHAKE256-DRBG}};
\draw[seed] (qec.east) -- (hrng.west);
\draw[seed] (hrng.east) -- (drbg.west);
\draw[thick, violet!75, dashed, Stealth-Stealth] (drbg) -- (keccak);
\node[font=\tiny, text=violet!75, anchor=west] at (1.12,-2.5){shares Keccak};
\draw[seed] ([xshift=-10mm]drbg.north) |- (-1.05,-2.55) |- (kyber.west);
\node[font=\tiny, text=violet!75, align=right, anchor=east] at (-1.22,0.05)
  {internal seed $\to$ KeyGen\\(ML-KEM / ML-DSA);\\never crosses PCIe};

\begin{scope}[on background layer]
  \node[draw=black!55, dashed, rounded corners=5pt, inner sep=9pt, fill=blue!2,
        fit=(xdma)(kyber)(hsm)(dil)(keccak)(mem)(hrng)(drbg)] (die) {};
\end{scope}
\node[font=\tiny\itshape, text=black!60, anchor=north east, inner sep=0pt]
  at ([shift={(-0.18,-0.16)}]die.north east) {FPGA die boundary};
\end{tikzpicture}
\caption{Top-level block diagram.  Green (dashed)~= shared datapath
multiplexed at runtime among the scheme FSMs, custody, and DRBG.  Orange
(bold)~= key custody on the secret-key path.  Violet~= entropy subsystem
(off-die Am-241 QEC $+$ on-die ring-oscillator TRNG $\to$ SHA3-256 $\to$
SHAKE256-DRBG, which time-shares the Keccak).  The host, outside the die
boundary, is untrusted for key confidentiality.}
\label{fig:blockdiagram}
\end{figure*}

As Figure~\ref{fig:blockdiagram} shows, the host communicates over
AXI4-Lite (BAR0 control/status registers) and AXI4-Stream
(host-to-card/card-to-host, H2C/C2H, bulk data) via Xilinx XDMA IP.  Clock domains: 100\,MHz core (\reg{pqc\_top\_clk}), 250\,MHz XDMA AXI
(\reg{axi\_aclk}), 500\,MHz PCIe transceiver (\reg{userclk1})
(Section~\ref{subsec:board}).

\textbf{Shared NTT/INTT datapath.}  A single 5-stage pipelined butterfly
(5-cycle latency, one butterfly per cycle after fill) iterates all 7
(ML-KEM) or 8 (ML-DSA) NTT layers; twiddle ROMs hold both moduli's
tables with the active modulus selected per transform, and the twiddle
multiply is reduced by a two-cycle Montgomery~\cite{montgomery1985}
step.  NTT-domain multiplication
instead uses Barrett reduction~\cite{barrett1987}---$m =
5039$, $k = 24$ for ML-KEM ($q = 3329$) and $m = 8\,396\,807$, $k = 46$
for ML-DSA ($q = 8\,380\,417$)---keeping coefficients in the normal
domain: a degree-one \emph{base multiply} for ML-KEM's incomplete NTT,
a \emph{pointwise} multiply for ML-DSA's complete NTT.

\textbf{Shared Keccak core.}  A single Keccak-f[1600] unit serves both
control FSMs and the key-custody keywrap/deterministic random-bit
generator (DRBG) through a registered
two-master mux, whose one extra command-bus cycle, invisible to
direct-wired unit testbenches, caused a wrong-tag failure in an intermediate custody-era build
(v79c),
caught only on hardware.

\textbf{On-die entropy (HRNG).}  A hardware RNG
(\reg{qec\_aggregator}) supplies all internal randomness.  Four off-die
Am-241 quantum-entropy chip sites (QEC~\cite{park2020qec}; one selected
at runtime via \reg{HRNG\_CTRL[8:7]}), sampled as pulse-arrival times,
are XOR-combined with an on-die
ring-oscillator TRNG, then conditioned by a SHA3-256 core
\emph{private} to the subsystem.  The conditioned
256-bit words seed a SHAKE256-DRBG that has no Keccak core of its own:
it \emph{time-shares} the single PQC Keccak through a bypass mux under
hardware mutual exclusion---the same consolidation reflex that dropped
a redundant ${\approx}1.5$\,k-LUT Keccak during timing closure.  Per-channel SP~800-90B health tests (repetition-count,
adaptive-proportion) report through \reg{HRNG\_STATUS} but are advisory
in the deployed build; a sparse Am-241 pulse train trips them on a
healthy chip.  With \reg{HRNG\_CTRL[9]} set, ML-KEM and ML-DSA KeyGen
draw their 32-byte seed from the DRBG (the internal-seed path,
Section~\ref{subsec:lineage}), so the seed never crosses PCIe---unlike
prior PQC accelerators, which receive
seeds or keys from the host (Section~\ref{sec:related}).
Entropy-validation status: H7 (Section~\ref{sec:threats}).

\textbf{Hardware key custody (summary).}  The FPGA die is the trust
boundary: private keys are generated, wrapped under
$\mathit{KEK} = \mathrm{SHAKE256}(\mathit{KEK\_SALT} \| \mathit{DeviceDNA})$,
and used entirely on-die.  Only public material crosses PCIe: the
public key and a wrapped blob at provisioning, the re-presented blob
plus the digest or ciphertext at sign/decapsulate time.  Private keys never leave the die.
Internal-seed (zero-knowledge) provisioning is
also supported; the full custody protocol, trust model, and security
analysis are out of scope (Section~\ref{sec:conclusion}).

\subsection{Resource Utilization}
\label{subsec:resources}

The FIPS-final
conversion (Section~\ref{subsec:lineage}) cost ${\approx}4.2$\,k LUTs
and ${\approx}370$ registers over the v87 fit (${\approx}4.5$\,k and
${\approx}400$ over v88), raising slice occupancy from 96.9\,\% (v87)
to 98.5\,\% (effectively full; Table~\ref{tab:resources}) with
LUT-as-memory, BRAM, and DSP counts unchanged.
The intermediate v88 build is datapath-identical to v87 and fits at
${\approx}97.3$\,\% slice / ${\approx}82.8$\,\% LUT, \reg{userclk1}
$+0.042$\,ns.

\begin{table}[t]
\caption{Resource utilization, shipped v89 (\reg{f52fb148}),
         XC7K160T-FFG676-2, Vivado 2024.2 post-route.}
\label{tab:resources}
\centering
\begin{tabular}{lrrr}
\toprule
Resource          & Used    & Available & Utilization \\
\midrule
Slices            & 24{,}975 & 25{,}350  & 98.5\,\%    \\
Slice LUTs        & 88{,}478 & 101{,}400 & 87.3\,\%    \\
Slice registers   & 61{,}749 & 202{,}800 & 30.4\,\%    \\
BRAM-36Kb tiles   & 171.5   & 325        & 52.8\,\%    \\
DSP48E1           & 39      & 600        &  6.5\,\%    \\
\bottomrule
\end{tabular}
\end{table}

\textbf{Timing.}  All clock domains close with no failing endpoints
(0 of 178{,}247).  \reg{userclk1} (500\,MHz) \emph{sets} the global worst
setup slack at $+0.049$\,ns, with \reg{userclk2} (250\,MHz, \reg{axi\_aclk}) at
$+0.156$\,ns, \reg{pqc\_top\_clk} (100\,MHz) at
$+0.362$\,ns, and worst hold slack $+0.015$\,ns.

\textbf{Power.}  Total on-chip power is $5.55$\,W (dynamic $5.39$\,W,
static $0.16$\,W), of which the PQC core (\reg{u\_pqc}) draws $1.62$\,W
and the PCIe/XDMA subsystem (\reg{u\_xdma}) $3.33$\,W.

\subsection{Throughput and Latency}
\label{subsec:throughput}

Table~\ref{tab:throughput} reports best-case sustained throughput,
measured on the deployed board with the C host
library (\reg{bench\_throughput}~\texttt{--profile}), single-stream
back-to-back, byte-exact outputs, at the \emph{peak} fast-path level
(reset and drain elided where safe; H5, Section~\ref{subsec:hostproto}).
Peak-level compute share ranges from
transport-dominated ML-KEM KeyGen to largely compute-bound Decapsulation
and Sign.

\begin{table*}[t]
\caption{System throughput and latency, shipped v89,
         host-measured over PCIe at the peak fast-path level
         (integrated-system numbers, not core-cycle records).  Custody
         latencies were measured on the v88 build.}
\label{tab:throughput}
\centering
\begin{tabular}{llrr}
\toprule
Scheme    & Operation              & Throughput (ops/s) & Mean latency (ms) \\
\midrule
\multirow{3}{*}{ML-KEM-768}
          & KeyGen                 & 2667 & 0.375 \\
          & Encaps                 & 1857 & 0.538 \\
          & Decaps                 & 1377 & 0.726 \\
\midrule
\multirow{3}{*}{ML-DSA-65}
          & KeyGen                 &  768 & 1.303 \\
          & Sign                   &  480 & 2.083 \\
          & Verify                 &  786 & 1.273 \\
\midrule
\multirow{4}{*}{Custody latency}
          & On-chip wrap           & \multicolumn{2}{l}{${\approx}7$\,ms (provisioning, host-measured)} \\
          & On-chip unwrap         & \multicolumn{2}{l}{${\approx}0.6$\,ms (per sign-from-vault call)} \\
          & Sign from vault        & \multicolumn{2}{l}{17--21\,ms (unwrap $+$ sign $+$ blob transfer, reset/drain)} \\
          & Decapsulate from vault & \multicolumn{2}{l}{1--2\,ms (incl.\ unwrap + decaps)} \\
\bottomrule
\end{tabular}
\end{table*}

\textbf{Performance context.}  Optimized ML-KEM/ML-DSA AVX2 software on
modern x86 reaches per-operation latencies of order 0.1\,ms or below
(reject-dependent ML-DSA signing, a few tenths of a
millisecond)~\cite{kyber2018,dilithium2018}, under the PCIe round-trip
floor.  The project's week-1 design target of $\geq 1{,}000$
encapsulations/s is met with headroom; ML-KEM KeyGen and Decaps also clear
1{,}000\,ops/s.

\textbf{Core-cycle latency.}  ML-KEM Decapsulation takes ${\approx}67{,}550$~cycles on
the full Fujisaki--Okamoto re-encryption path (67,535--67,564 across
test vectors in RTL simulation of the shipped datapath); at 100\,MHz
that is ${\approx}0.68$\,ms of the 0.726\,ms measured mean.  ML-DSA
Signing is rejection-dependent at ${\approx}0.4$--$0.6$~M cycles;
Table~\ref{tab:throughput}'s lower 2.083\,ms (${\approx}0.21$\,M cycles)
reflects the benchmark's fixed low-rejection message.  A forward NTT
takes 907 (ML-KEM) or 1035 (ML-DSA) cycles per polynomial.

\subsection{BUG-IDE-037: The Methodology's Key Catch}
\label{subsec:bugide037}

\textbf{Detection and symptom.}  A later seed-driven randomized
mixed-operation soak (${\approx}2{,}200$-vector pool) caught a
verification-failing signature for one message class
(EXP-20260615-001); the preceding ${\approx}84{,}000$-check soak
(EXP-20260614-001) passed clean, prompting episode~1 of
Section~\ref{subsec:wrongepisodes}; a constant SHA-256 hash across
reruns proved the defect deterministic, not transient or thermal.

\textbf{Root cause.}  In \texttt{dilithium\_sign.v}, the
rejection-sampling norm check reads coefficients from the main BRAM with
2-cycle latency, but the $\|\mathbf{z}\|$, $\|\mathbf{r_0}\|$, and
$\|c \cdot t_0\|$ scan counts covered \emph{exactly} the polynomial, so
the last one or two coefficients arrived \emph{after} the accept/reject
decision.  The failing vector (vec76, reject-loop
iteration~5) hit this blind spot: its sole norm-violating coefficient,
$r_0[\mathit{poly}[5]][\mathit{coeff}[0]]$, was last in scan order; a spec-correct signer rejects the candidate
and first accepts at iteration~8.

\textbf{Fix and validation.}  The fix extends each scan by the BRAM
latency margin: $\mathbf{z}$ $1279 \to 1281$ (+2, wrapping),
$\mathbf{r_0}$ $1535 \to 1537$ (+2, wrapping), and $c \cdot t_0$
$1535 \to 1536$ (+1, linear, plus a garbage-tail guard).  On hardware the
failing vector re-tested 20/20 byte-exact and the 7/7 KAT regression
passed.

\textbf{Detectability.}  Every affected signature \emph{fails}
verification (leakage caveats: H8, Section~\ref{sec:threats}).
KAT vectors cannot trigger it: they never sample the message class that
drives the rejection loop to the blind-spot iteration.

\textbf{Reliability closure.}  An 8-hour soak
(\reg{c\_exhaustive\_8h\_soak\_hw}) ran 28,801.6~s, completing 779,945
checks with zero failures across ML-DSA provision (72,625),
sign-from-vault$\to$Verify on data-dependent messages (301,343), Verify
tamper-reject (37,738), ML-KEM encaps$\to$decaps-from-vault (295,673),
and zeroize durability (72,562); ML-KEM provision (4) is thin coverage,
a stated caveat.  A separate
C-library soak added 186,432 checks, also with zero failures.

\subsection{Comparison with Prior Work}
\label{subsec:sota}

All comparator figures in Table~\ref{tab:comparison} were verified
against primary sources (IACR ePrint,
TCHES/PeerJ/arXiv open-access versions).

\textbf{Why A0's latency is larger.}  A0 is
roughly an order of
magnitude slower per operation for three compounding reasons.  (i)~\emph{Measurement scope}: \emph{end-to-end
system} latency over PCIe versus the
prior rows' accelerator-core latency.  (ii)~\emph{Clock}: A0's PQC core runs at
100\,MHz on a Kintex-7 versus 270--375\,MHz UltraScale+ parts.
(iii)~\emph{Microarchitecture}: A0 \emph{serializes} the NTT through
a single time-shared 5-stage butterfly
(Section~\ref{subsec:architecture}) to fit both schemes \emph{plus} key custody in a
nearly full XC7K160T, so even in
clock-independent core cycles it spends ${\sim}6$--$12\times$ more than
the parallel comparators (ML-KEM Decaps ${\approx}67{,}550$ vs.\ ${\approx}5.5$--$11$\,k).  With AVX2 software below the
PCIe round-trip floor, a PCIe
accelerator cannot win on raw throughput \emph{by construction}; A0's
value is \emph{CPU offload and hardware key custody}, which no compared core provides.

\begin{table*}[t]
\caption{Selected unified ML-KEM$+$ML-DSA FPGA accelerators, per-operation
  latency in $\mu$s.  Lettered marks refer to the notes beneath the table.
  Row~A0 (this work) is \emph{end-to-end system} latency; the FPGA comparator
  rows are \emph{accelerator-core} latency at ${\approx}2.7$--$3.75\times$
  higher clocks, so the two are \emph{not directly comparable}.}
\label{tab:comparison}
\centering
\small
\setlength{\tabcolsep}{6pt}
\renewcommand{\arraystretch}{1.2}
\begin{tabular}{@{}llccrrrcc@{}}
\toprule
\textbf{\#} & \textbf{Work} & \textbf{Device} & \textbf{$F_{\max}$}
  & \textbf{LUT} & \textbf{DSP} & \textbf{BRAM}
  & \textbf{ML-KEM-768} & \textbf{ML-DSA-65} \\
  &  &  & \textbf{(MHz)} &  &  &
  & \textit{KG\,/\,Enc\,/\,Dec} & \textit{KG\,/\,Sign\,/\,Vrf} \\
\midrule
\multicolumn{9}{@{}l}{\textit{Unified Kyber $+$ Dilithium accelerators
  (closest comparators)}} \\[2pt]
A0 & \textbf{This work (v89)} & Kintex-7 XC7K160T & 100\,/\,500$^{a}$
   & 88,478$^{b}$ & 39$^{b}$ & 171.5$^{b}$
   & 375\,/\,538\,/\,726$^{i}$ & 1303\,/\,2083\,/\,1273$^{i}$ \\
A1 & KaLi~\cite{aikata2023kali} & ZCU102$^{c}$ & 270
   & 23,277 & 4 & 24
   & 23.2\,/\,29.11\,/\,41.82 & 87.5\,/\,179.9\,/\,96.8$^{d}$ \\
A2 & Dobias et al.~\cite{dobias2025unified} & ZCU102$^{c}$ & 375
   & 17,138 & 4 & 12.5
   & 9.4\,/\,10.9\,/\,14.8 & 41.7\,/\,88.5\,/\,44.6 \\
A3 & KiD~\cite{mandal2023kid} & ZU+\,/\,A7\,/\,Z7$^{e}$ & 294--342
   & 2,893--5,909 & 4--8 & 4.5--5.5
   & \multicolumn{2}{c}{\textit{NTT multiplier unit only}$^{f}$} \\
A4 & Wang et al.~\cite{wang2024codesign} & PolarFire SoC$^{g}$ & ---
   & ${<}5\%$ dev. & --- & ---
   & 3--5$\times$ SW speedup & 3--5$\times$ SW speedup \\
A5 & Sapphire~\cite{banerjee2019sapphire} & 40\,nm ASIC$^{h}$ & 72
   & 106\,kGE & --- & ---
   & \multicolumn{2}{c}{\textit{configurable multi-scheme}$^{h}$} \\
\bottomrule
\end{tabular}

\smallskip
\begin{minipage}{\textwidth}
\footnotesize
$^{a}$\,100\,MHz core clock, 500\,MHz PCIe transceiver clock.\quad
$^{b}$\,Device utilization: LUT 87.3\%, DSP 6.5\%, BRAM 52.8\%.\quad
$^{c}$\,ZCU102 is Zynq UltraScale+.\quad
$^{d}$\,Best-case figures.\quad
$^{e}$\,Zynq UltraScale+\,/\,Artix-7\,/\,Zynq-7.\quad
$^{f}$\,NTT multiplier unit only, not a full KEM or signature core;
conflict-free dual-port.\quad
$^{g}$\,PolarFire SoC (RISC-V); resource figure is device fraction.\quad
$^{h}$\,ASIC, not FPGA; area in kGE; configurable multi-scheme
(Kyber/Dilithium-class, RISC-V\,+\,NTT\,+\,\mbox{SHA-3}).\quad
$^{i}$\,End-to-end \emph{system} latency (host\,+\,PCIe
Gen2\,$\times$\,8\,+\,DMA), v89 measured at the peak fast path; the default
correctness-first path adds ${\approx}1$\,ms of inter-operation reset and
drain.  Includes on-chip key custody (DNA-rooted KEK, wrap,
sign/decaps-from-vault).  Prior-work rows are \emph{accelerator-core}
latency and are not directly comparable, since clock rates, host
interfaces, and custody overhead all differ.  Core-cycle anchors and the Sign low-rejection caveat:
Section~\ref{subsec:throughput}.
\end{minipage}
\end{table*}

\section{The PQC--PCIe Board Implementation}
\label{sec:pcie}

The artifact of Section~\ref{sec:artifact} is a deployed PCIe card with
a complete host stack.  Most of the hardware difficulty sits here: the
50\% integration/bring-up bucket (Table~\ref{tab:successbucket}) and
much of the 2026-05 dip to 41\% (Section~\ref{subsec:overtime}) trace
to Section~\ref{subsec:bringup}.

\subsection{Board and PCIe Subsystem}
\label{subsec:board}

The board (EYL ``PQC-HSM K160,'' model K160-PCIe, a working product
designation) carries the XC7K160T-FFG676-2 behind a PCIe edge connector, a Micron MT25QL128 SPI configuration flash, and the
Am-241 quantum-entropy subsystem of Section~\ref{subsec:architecture}.  Deployed builds draw
${\approx}5.4$--$5.6$\,W on-chip (${\approx}1.5$--$1.65$\,W in the PQC
core, the rest mostly PCIe/XDMA and transceivers).

Host attachment uses the AMD/Xilinx XDMA endpoint~\cite{xilinxpg195} at
Gen2 5.0\,GT/s $\times$8 (${\approx}3.0$\,GB/s modeled
effective DMA),
which exposes a 64-kB BAR0 window of 32-bit AXI4-Lite registers plus
one H2C and one C2H AXI4-Stream channel (128-bit at 250\,MHz),
width-adapted to the wrapper's 8-bit interface across three
clock-domain crossings into the 100\,MHz core domain.  The
XDMA-internal 500\,MHz \reg{userclk1}, chronically the tightest domain (the
timing-closure row of Table~\ref{tab:bugtax}), closed between
$+0.041$ and $+0.049$\,ns.

Reset is asymmetric: the BAR0 soft-reset (\reg{0x08}, added in
v29) resets all PQC sub-FSMs via a self-clearing 16-cycle pulse
(${\approx}50$--$100\,\mu$s host-observed, including the re-arm
poll), but the
XDMA DMA engines and H2C CDC FIFO reset only on PCIe Function-Level
Reset (FLR), and everything on the FLR-only side eventually produced a bug.  The physical workflow
(Vivado~Lab~2023.2 flashing, a cold power-cycle after every flash, an
XDMA kernel module rebuilt against the running kernel) was learned
failure by failure and encoded into the persistent memory stack
(Section~\ref{subsec:context}).

\subsection{Host--Device Protocol}
\label{subsec:hostproto}

Control uses a compact bank of BAR0 registers (\reg{CTRL},
\reg{STATUS}, \reg{SOFT\_RESET}, the \reg{HRNG\_*}
entropy and \reg{HSM\_*} custody banks, and Device-DNA readback; full
map in the released manual).  Dispatch is deliberately minimal: the host writes a 3-bit opcode to
\reg{CTRL}, streams input over H2C, and reads the fixed-length result
over C2H, with no framing headers or length fields (\reg{tlast}
delimits); the C2H read returning its expected byte count \emph{is}
the completion signal, since \reg{STATUS.done} is a one-cycle pulse no
PCIe-latency poll can catch (interrupts are unused), and
\reg{STATUS.busy} is busy-polled only as the re-arm gate.  Wire sizes (v89 FIPS-final, H2C\,$\to$\,C2H, bytes): ML-KEM
KeyGen 32\,$\to$\,1184, Encaps 1216\,$\to$\,1120, Decaps
3488\,$\to$\,32; ML-DSA KeyGen 32\,$\to$\,1952, Sign
4064\,$\to$\,3309, Verify 5293\,$\to$\,0.  In internal-seed mode
(\reg{HRNG\_CTRL[9]}) KeyGen seed bytes drop to zero and custody provisioning returns a
wrapped blob (nonce $+$ tag $+$ ciphertext: 4096~B for ML-DSA-65,
2464~B for ML-KEM-768).  The direct Sign/Decaps formats carry the
plaintext secret key over H2C (the compute/migration path);
custody sign/decaps-from-vault instead re-presents the host-held,
on-die-wrapped blob over BAR0 vault writes on each call and unwraps it
on-die (the blob-transfer term of Table~\ref{tab:throughput}).  The one exception to C2H result readback, ML-DSA Verify, reports
its result in \reg{STATUS[2]} \reg{dil\_valid}, a late discovery.

The protocol also carries five \emph{operational
disciplines}, none in any component datasheet---each a distilled,
logged silicon failure, protocol step~P5 made visible:

\begin{enumerate}[nosep, label=\textbf{D\arabic*.}, leftmargin=*]
  \item \textbf{Soft-reset + \reg{DVF\_STATE} idle poll before every
    Verify} (\reg{DVF\_STATE} taps the Verify FSM's live state over
    BAR0): else the second back-to-back Verify returns invalid
    \emph{deterministically} (200/200; BUG-IDE-036,
    Section~\ref{subsec:wrongepisodes}).
  \item \textbf{Inter-operation settle}: consecutive ML-DSA operations
    need ${\geq}1$\,ms separation.  The fast path uses a busy-poll
    re-arm gate plus an unconditional 0.5\,ms residue drain.
  \item \textbf{Warm-up burn-op at open}: a ${\approx}1$\,kB tail in
    the XDMA C2H \emph{engine} survives
    \texttt{close()}$\to$\texttt{open()} and soft-reset, shifting the
    next process's first result by one byte; one throwaway KeyGen
    flushes it.
  \item \textbf{FLR as transport recovery}: a wedged H2C path,
    unreachable by any drain or soft-reset, recovers by scripted PCIe FLR
    in ${\approx}6$\,s, no power-cycle.
  \item \textbf{Single-opener serialization}: the XDMA device admits one
    opener (\texttt{EBUSY}).  Multi-client use goes through the
    \reg{pqcd} daemon (Section~\ref{subsec:hoststack}).
\end{enumerate}

\subsection{The Bring-Up Record: Where the 50\% Bucket Was Earned}
\label{subsec:bringup}

The PCIe bring-up compressed into four weeks (bitstreams
v11--v58b): first Gen2~$\times$8 builds on 2026-05-04 (v11), scripted
flash on 05-17 (v13), first end-to-end
PCIe operation (ML-KEM KeyGen, v18) and first byte-exact result (v22)
on 05-21 after fixing X-initialized \reg{\$readmemh} ROMs and the
Vivado multi-write BRAM dissolution, and all
six operations FSM-complete on 05-22 (v29).  Byte-exactness climbed
4/6, 5/6, 6/6 across 05-26/-27/-28 (the last a host-side fix, no
reflash) and reached 6/6 at full back-to-back rate on 05-29
(mitigated host-side; D1).

Three failure families dominate the record (Table~\ref{tab:bugtax}):

\textbf{(a) Stream-handshake races in RTL.}  Combinational request
lines into register-derived readiness, one-cycle pulses against
back-pressure, target-switch first-byte duplication, and compute-gap
over-fetch all passed iverilog and failed on silicon; each fix was a
registered request, a sticky flag, or a one-deep skid buffer.

\textbf{(b) Host-vs-RTL misdiagnosis.}  The costliest family, and the
observability problem in its purest form: host, driver, DMA, and RTL
produce identical boundary symptoms, so byte-exactness locates a
failure without attributing it.  Three ``RTL bugs'' were host code: an
abandoned reader thread silently consuming device bytes (BUG-IDE-031);
a C2H drain that silently no-ops off the main thread (BUG-IDE-033);
and BUG-IDE-034, the host waiting for Verify result bytes the RTL
never emits on C2H---masking two RTL repair attempts (v49, v51) until
the register-level read was found on 05-27, its resolution exposing a
further host-side \reg{tr}-encoding mismatch (BUG-IDE-035).  The
inverse occurred once: the back-to-back defect (BUG-IDE-036, D1)
presented as a driver flake but is RTL-behavioral.  Attribution came
only from differential A/B on live hardware (direct vs.\ threaded, C
vs.\ Python, before vs.\ after FLR), distilled into standing rules: a
failure that follows the host-side variable indicts the host, and no
RTL hunt opens before a clean re-run.  An X-clean full-module
simulation, cheap via the golden-reference spine
(Section~\ref{subsec:goldenspine}), exonerated the single-operation
Verify datapath as correct since v51.

\textbf{(c) Transport-layer defects.}  XDMA C2H short-completions,
AXI-Stream padding misreads, and a
${\approx}$1-in-20 mixed-operation H2C wedge (\texttt{errno~512},
FLR-only) round out the family; the wedge was self-healed by the soak
harness for weeks until root-caused
(BUG-IDE-021).  Its fix took a
500-iteration A/B from 24 wedges to one, then zero across a
1{,}500-iteration confirmation, 12{,}825/12{,}825 byte-exact.

The methodology reading: almost nothing here is cryptographic---it is
boundary engineering, exactly where Table~\ref{tab:successbucket}
shows coin-flip reliability (Section~\ref{subsec:hitldata}).

\subsection{Host Software Stack}
\label{subsec:hoststack}

Four host surfaces ship with the board, all implementing the protocol and disciplines above.  A Python library
(\reg{pqc\_pcie\_host.py}) covers all six operations plus custody, D3
warm-up by default.  A dependency-free C99 library (\reg{libpqchsm}) is protocol-exact
against it and is the Table~\ref{tab:throughput} bench client.  The \reg{pqcd} daemon serializes
clients over the single-opener device (D5) and feeds a reference
OpenSSL-3 provider that signed a TLS-1.3 CertificateVerify
through hardware custody.  The demo/validation suites add the
\reg{check\_all\_six} byte-exact gate and the self-healing soak harness
of Section~\ref{subsec:bugide037}.  The
redundancy is deliberate: C/Python divergences repeatedly localized
transport bugs neither stack could attribute alone
(Section~\ref{subsec:bugtaxonomy}).

\subsection{Productization and Bitstream Lineage Beyond the Snapshot}
\label{subsec:lineage}

\textbf{Release v1.0} (first cut 2026-06-21 on bitstream v87) packaged
the bitstream, host libraries, daemon, suites, and manual.
\textbf{Release v1.1} (2026-07-08) ships bitstream v89, converting both
schemes from round-3-derived encodings to the final published
standards: FIPS-203/204 key-generation seed derivations, 4-bit
\reg{SimpleBitPack} $w_1$ encoding, a 48-byte challenge, a signature
grown from 3293 to 3309 bytes (the only wire-visible length change),
and FIPS-204 domain-separated message hashing.  With v89, device
signatures are byte-identical to the FIPS-204 wire format, validated
against the \reg{dilithium-py} oracle, so conforming external
verifiers (e.g., liboqs~\cite{liboqs}, OpenSSL) accept them with no
host-side converter, and the device verifies external signatures
directly.  Release v1.1 was re-validated on the board: full suite
set and adversarial soak, zero failures on v89.

Every corpus-derived Section~\ref{sec:empirical} statistic is quoted
at the frozen 2026-06-26 snapshot (v88); the one flagged exception is
Section~\ref{subsec:tokens}'s measurement window (six experiments
beyond the freeze).  Release v1.1 postdates the
snapshot; built under the same logged methodology, it stands as
lineage, not analyzed corpus (scoping: H3).  Release v1.0 and its v88 update
(2026-06-24) preceded the freeze by days, and their logs are counted
inside the frozen corpus (the release/infra/PM bucket of
Table~\ref{tab:successbucket}).

\section{Related Work}
\label{sec:related}
\begin{table*}[t]
\caption{Closest prior works applying LLMs to PQC hardware; last
  column: the section substantiating each axis.}
\label{tab:pqcllm}
\centering
\small
\renewcommand{\arraystretch}{1.05}
\begin{tabular}{@{}lp{3.8cm}p{3.8cm}p{4.6cm}c@{}}
\toprule
\textbf{Axis} & \textbf{Liao et al.~\cite{liao2026llm4pqc} (ISQED'26)}
  & \textbf{LLM4PQC~\cite{perera2026llm4pqc} (DATE'26)}
  & \textbf{This work} & \textbf{Sec.} \\
\midrule
Scope & Compute kernels (FALCON) & PQC cores via HLS refactoring & Complete cryptosystem with host protocol, HSM, and bring-up & \ref{sec:artifact}, \ref{sec:pcie} \\
Deployment & FPGA implementation, kernel level & Synthesizable RTL & Real Kintex-7 silicon over PCIe Gen2~$\times$8 & \ref{sec:pcie} \\
Key custody & --- & --- & On-chip custody for both schemes (HRNG-seeded) & \ref{sec:artifact} \\
Interaction model & Human-in-the-loop prompt iteration & Feedback-driven, agentic & Agentic CLI with file/shell tools, subagents & \ref{sec:methodology} \\
Dataset & --- & --- & 232 structured experiment logs & \ref{sec:empirical} \\
\bottomrule
\end{tabular}
\end{table*}

\begin{table}[t]
\caption{Commercial PQC silicon (mid-2026).  FW~= PQC as firmware on
         a secure processor (no dedicated datapath).}
\label{tab:commercial}
\centering
\scriptsize
\setlength{\tabcolsep}{3pt}
\renewcommand{\arraystretch}{1.15}
\begin{tabular}{@{}p{2.5cm}p{1.55cm}p{2.55cm}p{1.3cm}@{}}
\toprule
Product & Class & PQC in silicon & Status \\
\midrule
ST ST54M~\cite{st54m2026} & mobile SE & ML-KEM, ML-DSA & sampling 2026 \\
Samsung S3SSE2A~\cite{samsung2026s3sse2a} & embedded SE & ML-DSA-65 only & sampling 2026 \\
SEALSQ QS7001~\cite{sealsq2025qs7001} & IoT SE & ML-KEM, ML-DSA & shipping 2025 \\
Microchip MEC175xB~\cite{microchip2025mec175xb} & embedded ctrl.\ (RoT) & ML-KEM, ML-DSA, LMS & sampling 2025 \\
Lattice MachXO5-NX TDQ~\cite{lattice2025machxo5} & secure-control FPGA & ML-KEM; LMS/XMSS bitstream auth & shipping 2025 \\
Caliptra / Adams Bridge~\cite{adamsbridge2026} & RoT IP (datacenter SoC) & ML-DSA-87 $+$ ML-KEM-1024, unified & RTL 2025 (IP-only) \\
IBM Crypto Express 8S~\cite{ibm4770} & PCIe HSM & none (Kyber / Dilithium in CCA FW) & shipping 2022 \\
Entrust nShield 5~\cite{entrust2025nshield} & PCIe / net.\ HSM & ML-KEM / ML-DSA / SLH-DSA;
  internal-FPGA accel.\ of
  undisclosed per-algorithm scope & FW 13.8, 2025 \\
\midrule
\textbf{This work} & \textbf{PCIe HSM accel.} & \textbf{ML-KEM-768$+$ML-DSA-65 full datapaths, on-die custody} & \textbf{deployed (v89)} \\
\bottomrule
\end{tabular}
\end{table}

\subsection{LLMs for RTL Generation and Hardware Benchmarking}
\label{subsec:related_llmgen}

Thakur et al.~\cite{thakur2023benchmark} introduced RTL-generation
benchmarks, VerilogEval~\cite{liu2023verilogeval} formalized pass@$k$
evaluation, and RTLCoder~\cite{liu2025rtlcoder} contributed an open
fine-tuned model and training set; ChipNeMo~\cite{liu2023chipnemo}
domain-adapts models, and Chip-Chat~\cite{blocklove2023chipchat}
carried an LLM-co-designed microcontroller through a SkyWater
130\,nm shuttle tapeout.  The artifacts are snippets, modules, or that shuttle core validated
at simulation, lint, bench demo, or expert rating; none reaches
hardened, in-service silicon with a longitudinal outcome record.

\subsection{Agentic and Feedback-Driven LLM Hardware Flows}
\label{subsec:related_agentic}

A second line closes a tool-in-the-loop cycle:
AutoChip~\cite{thakur2023autochip} iterates on testbench feedback,
RTLFixer~\cite{tsai2024rtlfixer} repairs syntax,
MEIC~\cite{xu2024meic} and VerilogCoder~\cite{ho2024verilogcoder}
add functional debugging and planning,
AssertLLM~\cite{fang2025assertllm} generates assertions, and
ChatEDA~\cite{he2024chateda} orchestrates the EDA toolchain on
benchmark designs (survey:~\cite{abdollahi2025scoping}).  None reaches
validated operation on deployed silicon.
QiMeng~\cite{zhang2025qimeng} spans RISC-V CPU spec-to-silicon but
stops at bring-up of a demonstrator taped out by its pre-LLM
pipeline.

\subsection{LLMs for Hardware Security}
\label{subsec:related_hwsec}

Ahmad et al.~\cite{ahmad2024hwsecbug} use LLMs to fix security
bugs in existing RTL rather than \emph{synthesize} cryptographic
hardware (survey:~\cite{akyash2024evolutionary}); to our knowledge, no published work carries LLM-driven design with a
cryptographic end-goal to validated operation on real silicon.

\subsection{Closest Prior Art: LLMs for PQC Hardware}
\label{subsec:related_llm4pqc}

Two 2026 works are closest: LLM4PQC~\cite{perera2026llm4pqc} and
Liao et al.~\cite{liao2026llm4pqc}, the latter reporting up to
$2.6\times$ kernel speedup over an HLS baseline.  Both stop short of validated operation, whereas ours continues
through byte-exact silicon and adversarial soak;
Table~\ref{tab:pqcllm} details the five-axis delta.

\subsection{ML-KEM/ML-DSA Hardware Accelerators}
\label{subsec:related_pqc}

FPGA acceleration of CRYSTALS-Kyber (ML-KEM,
FIPS~203~\cite{fips203,kyber2018}) spans a Pareto front from Xing and
Li's compact 2-DSP/3-BRAM Artix-7 design~\cite{xing2021kyber} to the
pipelined HPKA~\cite{huang2022hpka}; high-speed ML-DSA designs use
parallel butterflies and multiple Keccak cores ($2\times2$
NTT~\cite{beckwith2021dilithium}); area-leaner designs cover all
round-3 parameter sets~\cite{land2021dilithium}, with compact
variants at the far end of the area
axis~\cite{zhao2022dilithium,gupta2023lightweight}.

Most directly comparable are unified Kyber+Dilithium designs:
KaLi~\cite{aikata2023kali} (first, ZCU102, 270~MHz), Dobias et
al.~\cite{dobias2025unified} (most resource-efficient to date,
17,138~LUT at 375~MHz), KiD~\cite{mandal2023kid}, Wang et
al.~\cite{wang2024codesign}, Carril et al.~\cite{carril2024trets}
(HLS batch acceleration on a PCIe datacenter Alveo card), Beckwith et
al.~\cite{beckwith2023flexible}, and Truong et al.~\cite{truong2025unified}
(both FIPS-final standards, all phases, Zynq UltraScale+).  Ours is (1)~the only one, to our knowledge, on a mid-range Kintex-7;
(2)~integrated with on-chip key
custody \emph{and HRNG-seeded on-die key generation} for both schemes
(prior unified designs receive seeds and keys over the host bus); and (3)~measured end-to-end over PCIe including
host-driver and custody overhead, not at the accelerator boundary.  PUF-Dilithium~\cite{aghapour2025pufdilithium} also targets on-chip
key custody but stops at PUF-based seed protection of ARM software.  A complementary
line masks datapaths (among unified
designs, Beckwith et al.~\cite{beckwith2023flexible} report SCA
protection, and Dobias et al.~\cite{dobias2026twobirds} a
first-order-masked, TVLA-evaluated datapath); this work
deliberately makes no side-channel claim
(Section~\ref{sec:threats}, H6).

\textbf{Commercial PQC silicon.}  Table~\ref{tab:commercial} shows dedicated PQC silicon arriving in
2025--mid-2026.  Caliptra's open-source Adams Bridge accelerator
unifies ML-DSA-87 and ML-KEM-1024 as licensable
datacenter-SoC IP~\cite{adamsbridge2026}, the same consolidation this
design applies.  At HSM scale, most shipping products run PQC as
firmware on general-purpose secure processors~\cite{ibm4770}; the
closest analogue is Entrust's nShield~5, a classical HSM
field-upgraded with internal-FPGA
PQC acceleration~\cite{entrust2025nshield}.
To our knowledge, no commercial product yet documents full hardware
datapaths for both an ML-KEM and an ML-DSA parameter set with on-die
key custody at host-attached (PCIe) HSM scale; that is this
artifact's slot.

\section{Threats to Validity and Limitations}
\label{sec:threats}

We state nine honesty constraints---not hedges but integral to the
scientific claim.

\begin{enumerate}[label=\textbf{(H\arabic*)}, leftmargin=*, nosep]

  \item \textbf{$n{=}1$ case study / self-reported logs.}
    No human-only baseline was run: the productivity and cost figures
    of Section~\ref{subsec:advantages} are absolute, not comparative.
    Labels, lesson coding, and the per-log edit/iteration fields are
    self-recorded, incompletely covered (179 of 232; bounds in
    Section~\ref{subsec:hitldata}), and not independently
    adjudicated; a different segmentation would shift the percentages,
    not the coupling-gradient ordering.  Acceptance stringency also
    differs by bucket (operator judgment at low coupling, byte-exact
    at high; Section~\ref{subsec:bycategory}), so part of the
    gradient reflects adjudication.  Corpus completeness rests on
    contemporaneous logging (Section~\ref{subsec:logging});
    abandoned work scores as failure (Section~\ref{subsec:corpus}).

  \item \textbf{Single LLM family / single operator.}
    All sessions used Claude models and one developer; within this
    tightly coupled dyad, contributions are not separable, and
    generalizability to other LLMs, toolchains, operators, or task
    mixes is unknown---the bucket populations of
    Table~\ref{tab:successbucket} reflect this project's phase
    structure.

  \item \textbf{Trustworthiness from validation, not authorship.}
    Each RTL module is trusted because it passed byte-exact
    golden-reference comparison and, on the deployed v88 baseline, the
    779,945-check adversarial soak; the v89 release re-passed the
    full ladder, soak included (Section~\ref{subsec:lineage}).

  \item \textbf{Device binding, not at-rest key secrecy.}
    Given 7-series bitstream-encryption
    weaknesses~\cite{starbleed2020}, the bitstream-embedded
    $\mathit{KEK\_SALT}$ cannot be assumed secret against a physical
    adversary; a hardened key store is future work
    (Section~\ref{sec:conclusion}).

  \item \textbf{Integrated-system measurements, not a throughput
    record.}  Reported ops/s are measured at the peak fast-path level
    (the default correctness-first path adds ${\approx}1$\,ms of
    inter-operation reset and drain), and the ML-DSA Sign figure
    reflects the benchmark's fixed low-rejection message
    (Section~\ref{subsec:throughput}).

  \item \textbf{No side-channel resistance claims.}
    No masking is applied to the NTT, Keccak, or sampling datapaths
    (future work: Section~\ref{sec:conclusion}).

  \item \textbf{Not FIPS~140-3 validated.}
    The module is FIPS~203/204 byte-exact against the reference
    implementations (Section~\ref{subsec:lineage}) but has undergone
    neither Cryptographic Algorithm Validation Program (CAVP) testing
    nor FIPS~140-3 validation; likewise, while the QEC source's
    min-entropy is characterized per SP~800-90B
    methodology~\cite{park2020qec}, the deployed HRNG chain has no
    formal entropy-source validation---statistical batteries validate
    the conditioner output only.

  \item \textbf{BUG-IDE-037 is fail-detectable, not a forgery
    vulnerability.}
    The defect produces signatures that fail verification (fixed in
    v84); the rare pre-fix emissions
    are treated as a potential leakage surface, with no claim of
    security-neutrality.

  \item \textbf{Single-operation-at-a-time / XDMA single-opener.}
    ML-KEM and ML-DSA share the NTT and Keccak cores, so one
    operation executes at a time; the XDMA device is single-opener (D5).

\end{enumerate}

\section{Conclusion}
\label{sec:conclusion}

The standard acceptance test for lattice-signature hardware cannot
detect an entire class of defects: ML-DSA correctness is a property of
control paths while known-answer tests sample values; BUG-IDE-037 passed a full KAT regression on deployed
silicon.  We
resolved that blindness by replacing the gate rather than the author: acceptance
decided by a byte-exact golden-reference oracle plus randomized
adversarial soak---301,343 data-dependent signings among
779,945 checks with zero escapes on the deployed v88 baseline.  Nothing in that construction
depends on who wrote the RTL, which made the AI question
answerable.  Across 232 logged experiments an agentic LLM
assistant drove RTL through hardening at 71.6\% success, following a
\emph{hardware-coupling gradient} (77--85\% documentation/research
vs.\ 50--53\% synthesis and bring-up).  The
mechanism is observability, not difficulty: reasoning over readable
artifacts cannot recover a corrective signal that exists only on the
physical side of the boundary.  The slope of this $n{=}1$ study is a
hypothesis with a mechanism attached, not a calibrated constant.  Not in question is the
artifact: a unified ML-KEM-768 and ML-DSA-65
accelerator with on-chip key custody on one mid-range XC7K160T at
98.5\% slice occupancy and 49\,ps of margin at 500\,MHz, byte-exact
across all six FIPS operations (shipped v89, re-soaked, zero
failures).

\textbf{Future directions.}  (i)~A controlled multi-operator,
multi-model replication with baselines and independent outcome
adjudication beyond this $n{=}1$ case;
(ii)~genuine at-rest key-encryption-key secrecy via an eFUSE-protected
UltraScale+ (XCZU7EV) port, already first-silicon byte-exact across
all six operations under the same methodology;
(iii)~side-channel hardening by threshold implementation, where
LLM-assisted masked-gadget generation and RTL-level leakage
repair~\cite{srivastava2023scar} could lower its cost;
(iv)~higher-throughput multi-butterfly NTTs on larger devices, in
deployments where the core rather than the PCIe round trip bounds
throughput; (v)~the deferred custody-protocol specification and
analysis, plus FIPS~140-3 and CAVP validation for regulated
deployments; and (vi)~context condensation---distilled digests of the
slowly changing corpus to cut the context cost center directly
(Section~\ref{subsec:advantages}), in measured tension with the
stale-prose lesson of Section~\ref{subsec:wrongepisodes}.  All
inherit the same discipline: state the rung of
validation that earns the claim, and claim nothing above it.

\section*{Author Contributions and Acknowledgments}
J.~Park conducted the 232-experiment campaign as the single
operator and wrote the manuscript; E.~Kim, W.~Kim, and S.~Cho
(EYL Inc.) provided and supported the PQC--PCIe K160 board platform;
B.~Cha critically reviewed and revised the manuscript.  Anthropic
Claude models served as the LLM assistant under the human-gated
methodology of Section~\ref{sec:methodology}.

\bibliographystyle{IEEEtran}
\bibliography{refs}

\end{document}